\documentclass[aps,prd,groupedaddress,twocolumn,floatfix,nofootinbib,showpacs,showkeys]{revtex4-1}

\usepackage{graphicx}
\usepackage{dcolumn}
\usepackage{amsmath}
\usepackage{epstopdf}
\usepackage{graphicx}

\def\mnras{MNRAS}
\def\apj{Astroph. J.}
\def\apjl{Astroph. J. Lett.}

\def\aap{Astron. Astroph.}

\def\araa{Ann. Rev. Astron. Astroph.}
\def\nat{Nature}

\begin{document}

\title{Fast Rotating Neutron Stars with Realistic Nuclear Matter Equation of State}
\author{F.~Cipolletta,$^{1}$ C.~Cherubini,$^{2,3}$ S.~Filippi,$^{2,3}$ J.~A.~Rueda,$^{1,4,5}$, R.~Ruffini$^{1,4,5}$}
\email{cipo87@gmail.com; c.cherubini@unicampus.it;\\s.filippi@unicampus.it; jorge.rueda@icra.it; ruffini@icra.it}
\affiliation{$^1$Dipartimento di Fisica and ICRA, Sapienza Universit\`a di Roma, P.le Aldo Moro 5, I--00185 Rome, Italy}
\affiliation{$^2$Nonlinear Physics and Mathematical Modeling Lab, University Campus Bio-Medico of Rome, Via A.~del Portillo 21, I--00128 Rome, Italy}
\affiliation{$^3$International Center for Relativistic Astrophysics-ICRA, University Campus Bio-Medico of Rome, Via A.~del Portillo 21, I--00128 Rome, Italy}
\affiliation{$^4$ICRANet, Piazza della Repubblica 10, I--65122 Pescara, Italy}
\affiliation{$^5$ICRANet-Rio, Centro Brasileiro de Pesquisas F\'isicas, Rua Dr. Xavier Sigaud 150, Rio de Janeiro, RJ, 22290--180, Brazil }
\date{\today}

\date{\today}

\begin{abstract}
We construct equilibrium configurations of uniformly rotating neutron stars for selected relativistic mean-field nuclear matter equations of state (EOS). We compute in particular the gravitational mass ($M$), equatorial ($R_{\rm eq}$) and polar ($R_{\rm pol}$) radii, eccentricity, angular momentum ($J$), moment of inertia ($I$) and quadrupole moment ($M_2$) of neutron stars stable against mass-shedding and secular axisymmetric instability. By constructing the constant frequency sequence $f=716$~Hz of the fastest observed pulsar, PSR J1748--2446ad, and constraining it to be within the stability region, we obtain a lower mass bound for the pulsar, $M_{\rm min}=[1.2$--$1.4]~M_\odot$, for the EOS employed. Moreover we give a fitting formula relating the baryonic mass ($M_b$) and gravitational mass of non-rotating neutron stars, $M_b/M_\odot=M/M_\odot+(13/200)(M/M_\odot)^2$ [or $M/M_\odot=M_b/M_\odot-(1/20)(M_b/M_\odot)^2$], which is independent on the EOS. We also obtain a fitting formula, although not EOS independent, relating the gravitational mass and the angular momentum of neutron stars along the secular axisymmetric instability line for each EOS. We compute the maximum value of the dimensionless angular momentum, $a/M\equiv c J/(G M^2)$ (or ``Kerr parameter''), $(a/M)_{\rm max}\approx 0.7$, found to be also independent on the EOS. We compare and contrast then the quadrupole moment of rotating neutron stars with the one predicted by the Kerr exterior solution for the same values of mass and angular momentum. Finally we show that, although the mass quadrupole moment of realistic neutron stars never reaches the Kerr value, the latter is closely approached from above at the maximum mass value, as physically expected from the no-hair theorem. In particular the stiffer the EOS is, the closer the Kerr solution is approached. 
\end{abstract}



\maketitle

\section{Introduction}\label{sec:1}

Understanding the physics of neutron stars and being able to describe the spacetime in the interior and around such compact objects is one of the most important objectives for modern astrophysics, as from one side it could be an excellent test for the general relativity and on the other because the strong forces acting and the high level of densities reached cannot be tested in any place in the earth.

Although our actual incapability to describe these concepts exactly, some observational limits has already been determined, via simplifying assumptions (such as a spherical configuration in the X-ray binaries observation). Other general considerations on the nature of neutron stars and pulsars are often extracted in the literature from the use of fiducial structure parameters: a canonical neutron star (NS) of mass $M=1.4~M_\odot$, radius $R=10$~km, and moment of inertia $I=10^{45}$~g~cm$^2$ (see, e.g., \cite{caraveo14}, and references therein). Based on these parameters together with the so inferred surface magnetic field from the classic point-like magneto-dipole rotating model \cite{pacini67,gold68}, NSs have been traditionally classified according to the thought nature of the energy source powering their observed emission. Neutron stars are also thought to possibly participate in the most powerful explosions in the Universe, the gamma-ray bursts (GRBs), e.g. via NS mergers in the case of short GRBs (see, e.g, \cite{2014ApJ...787..150O}, and references therein) and hypercritical accretion processes leading to gravitational collapse to a black hole (BH) in the case of long GRBs associated to supernovae (see, e.g., \cite{2012ApJ...758L...7R, 2014ApJ...793L..36F}, and references therein).

There are still, however, many open issues regarding the above global picture of NSs both from the physics and astrophysics points of view (see, e.g., \cite{2013FrPhy...8..679H,2015arXiv150100005T}). On the other hand, our theoretical and observational knowledge on NSs has been largely increased in the intervening years from the first general relativistic description of a NS by Oppenheimer and Volkoff in 1939 \cite{1939PhRv...55..374O}. Eventually a more complex equation of state (EOS), interior structure, and consequently exterior gravitational field of non-rotating, slowly rotating, and fast rotating rotating stars have been acquired, and massive NSs of $\approx 2~M_\odot$, drastically constraining the nuclear EOS stiffness, have been observed \cite{antoniadis12, antoniadis13}. Thus, general conclusions based on fiducial parameters and corresponding observables, might be premature and a more exhausting exploration of the consequences of adopting different NS parameters appears to be necessary. 

For instance, as recently shown in \cite{2015ApJ...799...23B} for the high-magnetic field pulsar class, thought to be intermediate objects linking pulsars and magnetars, the understanding of their magnetic field values and their energy source needs the precise knowledge of the entire set of possible structure parameters of a NS and a self-consistent general relativistic description of the pulsar observables. Indeed, it was there shown how the magnetic field of a pulsar is overestimated, and the rotational energy underestimated, by the classic magnetic rotating dipole model and the use of fiducial NS parameters. 

The description of the rotational and thermal evolution, as well as the emitted radiation of isolated and accreting NS requires the knowledge of the structure properties and the corresponding exterior metric. For instance, we have recently compared and contrasted the cooling evolution of neutral NSs satisfying local charge neutrality and global charge neutrality \cite{2014PhRvC..90e5804D}. We have shown there that, owing to their different crust structure (mass and thickness) for a same value of the total mass, their thermal relaxation time (the time to form an isothermal core) can be very different and therefore the signatures of the structure of the NS might be accessed via early cooling observations.

There are recent numerical relativity's computations of the structure of uniformly rotating NSs, which have mainly focused on the existence of universal relations between e.g. the quadrupole moment, moment of inertia, and Love number of NSs (see, e.g., \cite{2012PhRvL.108w1104P,2014PhRvD..89l4013Y}, and \cite{2013Sci...341..365Y,2013PhRvD..88b3009Y}), considered in a slow rotation regime. Other works (e.g.~\cite{2014ApJ...781L...6D,2014PhRvL.112t1102C}) tried to recover these relation also in a full rotation regime, with a numerical method based on the work firstly implemented in \cite{1994ApJ...424..823C}.

In this work, through a full rotation approach (treated by numerical relativity's methods), we will focus on additional structure properties of uniformly rotating NSs relevant for astrophysical applications such as mass, polar and equatorial radii, eccentricity, angular momentum, angular velocity, moment of inertia, and quadrupole moment, for a selected sample of EOS (describing nuclear matter with relations of different stiffness) within relativistic mean-field nuclear theory, not analyzed in the set of EOS of previous works. 

This article is organized as follows. In section \ref{sec:2} we briefly review the axisymmetric system of Einstein's equations to be integrated for a given EOS which we describe in section \ref{sec:3}. The stability conditions (mass-shedding and secular instability) are outlined in section \ref{sec:4} and the mass-radius of rotating NSs is shown in section \ref{sec:5}. The eccentricity and the moment of inertia are shown in section \ref{sec:6} while the quadrupole moment is discussed in section \ref{sec:7}. We finally summarize and discuss our results in section \ref{sec:8}.

\section{Structure Equations}\label{sec:2}

We consider the equilibrium equations for a self-gravitating, rapidly rotating NS, within a fully general relativistic framework. We start with the stationary axisymmetric spacetime metric (see, e.g.,~\citep{2003LRR.....6....3S}):
\begin{equation}
\label{eq1}
ds^2 = -e^{2 \nu} dt^2 + e^{2 \psi} (d \phi - \omega dt)^2 + e^{2 \lambda} (dr^2 + r^2 d \theta^2),
\end{equation}
where $\nu$, $\psi$, $\omega$ and $\lambda$ depend only on variables $r$ and $\theta$. It is useful to introduce the variable $e^{\psi} =  r \sin(\theta) B e^{-\nu}$, being again $B = B(r,\theta)$. The above form of the metric is obtained under two assumptions: 1) the Killing vector fields are, one timelike $t^a$, and one relative to the axial symmetry, $\phi^a$; 2) the spacetime is asymptotically flat. Then, one can introduce \emph{quasi-isotropic coordinates}, which in the non-rotating limit they tend to isotropic ones.

Turning to the physical matter content in the NS interior, if one neglects sources of non-isotropic stresses, viscosity, and heat transport, then the energy-momentum tensor becomes the one of a perfect-fluid
\begin{equation}
\label{eq2}
T^{\alpha \beta} = (\varepsilon + P) u^\alpha u^\beta + P g^{\alpha \beta},
\end{equation}
where $\varepsilon$ and $P$ denote the energy density and pressure of the fluid, and $u^\alpha$ is the fluid 4-velocity. In terms of the two Killing vectors, 
\begin{equation}
u^\alpha =\frac{e^{- \nu}(t^\alpha + \Omega \phi^\alpha)}{\sqrt{1-v^2}},
\end{equation}
where $v$ is the fluid 3-velocity with respect to the local zero angular momentum observer (ZAMO), 
\begin{equation}
v = (\Omega - \omega) e^{\psi-\nu}
\end{equation}
being $\Omega \equiv u^\phi/u^t = d\phi/dt$ the angular velocity in the coordinate frame, equivalent to the one measured by an observer at rest at infinity. 

Thus, with the metric given by equation~(\ref{eq1}) and the energy-momentum tensor given by equation~(\ref{eq2}), one can write the field equations as (analogoulsly to Ref.~\cite{1976ApJ...204..200B} setting $\zeta=\lambda + \nu$):


\begin{eqnarray}
\label{eqA1a}
\nabla\cdot\left( B \nabla \nu  \right) &=& \frac{1}{2} r^2 \sin^2 \theta B^3 e^{-4\nu} \nabla \omega \cdot \nabla \omega \nonumber \\ 
&+& 4 \pi B e^{2\zeta - 2\nu} \left[ \frac{(\varepsilon+P)(1+v^2)}{1-v^2} + 2P \right],
\end{eqnarray}

\begin{eqnarray}
\label{eqA1b}
\nabla \cdot \left(  r^2 \sin^2 \theta B^3 e^{-4\nu} \nabla \omega \right) &=& -16 \pi r \sin \theta B^2 \nonumber\\
&\times& e^{2\zeta - 4\nu} \frac{(\varepsilon +P)v}{1-v^2},
\\
\label{eqA1c}
\nabla \cdot \left( r \sin(\theta) \nabla B \right) &=& 16 \pi r \sin \theta B e^{2\zeta - 2\nu} P,
\end{eqnarray}

\begin{widetext}

\begin{eqnarray}
\label{eqA1d}
{\zeta}_{,\mu} =& - & {\left\lbrace \left( 1-{\mu}^2 \right) {\left( 1+r \frac{B_{,r}}{B} \right)}^2 + {\left[ \mu - \left( 1-{\mu}^2 \right) \frac{B_{,r}}{B} \right]}^2 \right\rbrace}^{-1}  \Biggl[\frac{1}{2} B^{-1} \left\lbrace r^2 B_{,rr} - {\left[ \left( 1 -{\mu}^2 \right) B_{,\mu} 
                       \right]}_{,\mu} - 2\mu B_{,\mu} \right\rbrace  \nonumber \\
                       & \times & \left\lbrace - \mu + \left( 1-{\mu}^2 \right) \frac{B_{,\mu}}{B} \right\rbrace +  r\frac{B_{,r}}{B} \left[ \frac{1}{2} \mu + \mu r\frac{B_{,r}}{B} + \frac{1}{2} \left( 1-{\mu}^2 \right) \frac{B_{,\mu}}{B} \right] + \frac{3}{2} \frac{B_{,\mu}}{B} \left[ -{\mu}^2 + \mu \left( 1-{\mu}^2 \right) \frac{B_{,\mu}}{B} \right]\nonumber\\
                       & - & \left( 1-{\mu}^2 \right) r\frac{B_{,\mu r}}{B} \left( 1+r\frac{B_{,r}}{B} \right) - \mu r^2 {\left( {\nu}_{,r} \right)}^2 - 2\left( 1-{\mu}^2 \right) r{\nu}_{,\mu}{\nu}_{,r} + \mu \left( 1-{\mu}^2 \right){\left( {\nu}_{,\mu} \right)}^2 - 2\left( 1-{\mu}^2 \right) r^2 B^{-1} B_{,r} {\nu}_{,\mu} {\nu}_{,r}\nonumber \\
                       & + & \left( 1-{\mu}^2 \right) B^{-1} B_{,\mu} \left[ r^2 {\left( {\nu}_{,r} \right)}^2 - \left( 1-{\mu}^2 \right) {\left( {\nu}_{, \mu} \right)}^2 \right] + \left( 1-{\mu}^2 \right) B^2 e^{- 4\nu} \Bigl\lbrace \frac{1}{4} \mu r^4 {\left( {\omega}_{,r} \right)}^2 + \frac{1}{2} \left( 1-{\mu}^2 \right) r^3 {\omega}_{,\mu} {\omega}_{,r}\nonumber \\
                       & - & \frac{1}{4} \mu \left( 1-{\mu}^2 \right) r^2 {\left( {\omega}_{, \mu} \right)}^2 + \frac{1}{2} \left( 1-{\mu}^2 \right) r^4 B^{-1} B_{,r} {\omega}_{,\mu} {\omega}_{,r} - \frac{1}{4} \left( 1-{\mu}^2 \right) r^2B^{-1}B_{,\mu} \left[ r^2 {\left( {\omega}_{,r} \right)}^2 - \left( \-{\mu}^2 \right) {\left( {\omega}_{,\mu} \right)}^2 \right] \Bigr\rbrace \Biggr],
\end{eqnarray}

\end{widetext}
where, in the equation for ${\zeta}_{,\mu}$, we introduced $\mu\equiv \cos(\theta)$. 

The projection of the conservation of the energy-momentum tensor, normal to the 4-velocity, $ \left( \delta^{c}_{b} + u^{c} u_{b} \right) \nabla_{a} T^{ab} = 0 $, leads to the hydrostationary equilibrium equation:
\begin{equation}
\label{eq5}
P_{,i} + \left( \varepsilon + P \right) \left[ {\nu}_{,i} + \frac{1}{1-v^2} \left(- v v_{,i} + v^2 \frac{{\Omega}_{,i}}{\Omega - \omega}  \right) \right] = 0,
\end{equation}
where $i=1,2,3$ and, as usual, $A_{,i}\equiv \partial A/\partial x^i$.

For a barotropic equation of state (EOS), $P=P(\varepsilon)$, and in the case of uniform rotation which we adopt in this work, the above hydrostationary equilibrium equation has a first integral that can be written as
\begin{equation}
\label{eq6}
{\int}_0^{P} \frac{dP}{\varepsilon + P} - \ln (u^a {\nabla}_a t) =  {\nu}\vert_{\rm pole},
\end{equation}
where the constant of motion has been obtained, for instance, at the pole of the star (see, e.g.,~Ref.~\cite{2003LRR.....6....3S}).

\section{Equation of State}\label{sec:3}

To obtain a solution to the field equations, the matter EOS must be supplied. In general, a NS is composed of two regions, namely the core and the crust. The core, with densities overcoming the nuclear saturation value, $\rho_{\rm nuc}\approx 3\times 10^{14}$~g~cm$^{-3}$, is composed by a degenerate 
gas of baryons (e.g.~neutrons, protons, hyperons) and leptons (e.g.~electrons and muons). The crust, in its outer region ($\rho \leq \rho_{\rm drip}\approx 4.3\times 10^{11}$~g~cm$^{-3}$), is composed of ions and electrons, while in its inner region ($\rho_{\rm drip}<\rho<\rho_{\rm nuc}$), there is an additional component of free neutrons dripped out from nuclei. For the crust, we adopt the Baym-Pethick-Sutherland (BPS) EOS \cite{1971ApJ...170..299B}. For the core, we here adopt modern models based on relativistic mean-field (RMF) theory. Indeed, RMF models have become the most used ones in NS literature, being its success mainly owing to important properties such as Lorentz covariance, intrinsic inclusion of spin, a simple mechanism of saturation for nuclear matter, and being consistently relativistic, they do not violate causality (see, e.g., Ref.~\cite{2014PhRvC..89c5804R}). We adopt, as now becoming traditional, an extension of the original formulation of Boguta and Bodmer \cite{1977NuPhA.292..413B} in which nucleons interact via massive meson mediators of different nature providing the attractive long range (scalar $\sigma$) and repulsive short range (vector $\omega$) of the nuclear force, isospin and surface effects (vector $\rho$). Meson-meson interactions can be also present; for instance, in the version of Boguta and Bodmer \cite{1977NuPhA.292..413B} there is the presence of a self-interacting scalar field potential in form of a quartic polynom with adjustable coefficients. We consider here the possibility of including, in addition to such potential, vector-vector interactions of the $\omega$ meson. For a very recent and comprehensive analysis of the performance of several RMF models in the description of observed properties of ordinary nuclei, we refer the reader to Ref.~\cite{2014PhRvC..90e5203D}, and for a brief historical and chronological reconstruction of the developments of the RMF models, Ref.~\cite{2012NuPhA.883....1B}. 

Thus, we shall constrain ourselves to models in which the energy density and pressure are given by (in units with $\hbar = c =1$) \cite{2014PhRvC..90e5203D}:
\begin{widetext}

\begin{subequations}
\begin{eqnarray}
\label{eq21a}
\varepsilon &=& \frac{1}{2} m_{\sigma}^2 {\sigma}^2 + \frac{g_{\sigma 2}}{3} {\sigma}^3 + \frac{g_{\sigma 3}}{4} {\sigma}^4 - \frac{1}{2} m_{\omega}^2 {\omega}_0^2 - \frac{g_{\omega 3}}{4} (g_{\omega}^2 {\omega}_0^2)^2 - \frac{1}{2} m_{\rho}^2 \rho_{0}^2 + g_{\omega} {\omega}_0 n_{B} + \frac{g_{\rho}}{2} \rho_{0}  n_{3} + \sum_{i=n,p,e,\mu} \varepsilon_i,
\\
\label{eq21b}
P &=& -\frac{1}{2} m_{\sigma}^2 {\sigma}^2 - \frac{g_{\sigma 2}}{3} {\sigma}^3 - \frac{g_{\sigma 3}}{4} {\sigma}^4 + \frac{1}{2} m_{\omega}^2 {\omega}_0^2 + \frac{g_{\omega 3}}{4} (g_{\omega}^2 {\omega}_0^2)^2 + \frac{1}{2} m_{\rho}^2 \rho_{0}^2  + \sum_{i=n,p,e,\mu} P_i,
\end{eqnarray}
\end{subequations}

\end{widetext}
where $m_{\sigma,\omega,\rho}$ are the masses of the scalar and vector mesons, $g_{\sigma 2,3}$, $g_{\omega}$, $g_{\omega3}$ are coupling constants, $\sigma$, $\omega_0$, $\rho_0$ denotes the scalar meson and the time-component of the $\omega$ and $\rho$ vector mesons, respectively. The components $\varepsilon_i$ and $P_i$ for each kind of particle considered are
\begin{subequations}
\begin{eqnarray}
\label{eq22a}
\varepsilon_{n,p} &=& \frac{2}{(2\pi)^3} \int_0^{k_{n,p}^F} \sqrt{k^2 + (m_{n,p}^{\ast})^2} d^3k,
\\
\label{eq22b}
\varepsilon_{e,\mu} &=& \frac{2}{(2\pi)^3} \int_0^{k_{e,\mu}^F} \sqrt{k^2 + (m_{e,\mu})^2} d^3k,
%
\\
\label{eq23a}
P_{n,p} &=& \frac{1}{3} \frac{2}{(2\pi)^3} \int_0^{k_{n,p}^F} \frac{k^2}{\sqrt{k^2 + (m_{n,p}^{\ast})^2}} d^3k,
\\
\label{eq23b}
P_{e,\mu} &=& \frac{1}{3} \frac{2}{(2\pi)^3} \int_0^{k_{e,\mu}^F} \frac{k^2}{\sqrt{k^2 + (m_{e,\mu})^2}} d^3k,
\end{eqnarray}
\end{subequations}

where with $m_{i}^{\ast}$ is intended the effective mass of baryons.

The scalar, isospin, and baryon densities are given by, respectively,
\begin{subequations}
\begin{eqnarray}
\label{eq24b}
n_s &=& \frac{2}{(2\pi)^3} \sum_{i=n,p} \int_0^{k_{i}^F} \frac{m_{i}^{\ast}}{\sqrt{k^2 + (m_{i}^{\ast})^2}} d^3k,
\\
\label{eq24c}
n_3 &=& n_p - n_n,
\\
\label{eq24d}
n_B &=& n_p + n_n.
\end{eqnarray}
\end{subequations}
where, $n_i = (k_{i}^F)^3/(3\pi^2)$, are the particle number densities with $k_{i}^F$ the particle Fermi momenta.

The equations of motion of the meson fields within the RMF approximation are:
\begin{subequations}
\begin{eqnarray}
\label{eq25a}
m_{\sigma}^2 \sigma &=& g_{\sigma} n_s - g_{\sigma 2} {\sigma}^2 - g_{\sigma 3} {\sigma}^3,
\\
\label{eq25b}
m_{\omega}^2 {\omega}_0 &=& g_{\omega} n_B - g_{\omega 3} g_{\omega} (g_{\omega} {\omega}_0)^3, 
\\
\label{eq25c}
m_{\rho}^2 \rho_{0} &=& \frac{g_{\rho}}{2} n_3.
\end{eqnarray}
\end{subequations}

A barotropic EOS can be obtained iff additional closure relations are supplied. A first condition to be imposed is the request of the stability of matter against beta decay. The second closure equation that has been traditionally adopted is the condition of local charge neutrality of the system. The latter condition has been recently shown to be not fully consistent with the equilibrium equations in presence of multicomponent charged constituents such as protons and electrons (see \cite{2012NuPhA.883....1B}, and references therein). Instead, only global charge neutrality must be requested. The new system of equations, referred to as Einstein-Maxwell-Thomas-Fermi (EMTF) equations, introduce self-consistently the Coulomb interactions in addition to the strong, weak, and gravitational interactions within a full general relativity framework. It is worth to notice that in this case no perfect-like form of the total energy-momentum tensor is obtained since the presence of electromagnetic fields breaks the pressure isotropy. Static NSs fulfilling the EMTF equations were constructed in Ref.~\cite{2012NuPhA.883....1B}, and uniformly rotating configurations in the second-order Hartle approximation can be found in Ref.~\cite{2014NuPhA.921...33B}. To construct rotating NSs beyond the slow rotation regime, we take advantage of existing public numerical codes (e.g.~the RNS code, see section \ref{sec:5}) that solve the field equations without any limitation of the rotation rate of the star. However, an implementation of the equations and boundary conditions of the EMTF system within these codes is not yet available. Thus, as a first step, we adopt in here the condition of local charge neutrality, bearing in mind the necessity of a future implementation of the EMTF equations of equilibrium in fast rotation regime. 

\begin{figure}[!hbtp]
\includegraphics[width=\hsize,clip]{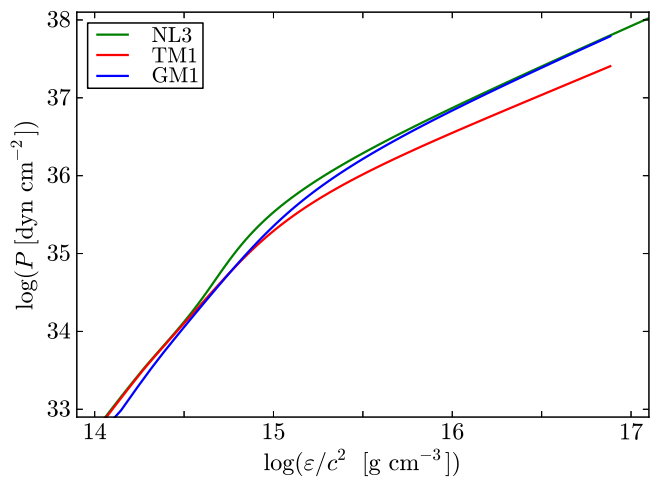}
\caption{\label{fig01}Pressure-energy density relation for the three EOS (TM1,GM1, and NL3) used in the present article.}
\end{figure}

With both the beta equilibrium and the local charge neutrality conditions, a numerical relation between the energy density and the pressure can be obtained. We here adopt the nuclear parameterizations (for the specific values of the coupling constants, particle, and meson field masses) NL3 \cite{1997PhRvC..55..540L}, TM1 \cite{1994NuPhA.579..557S}, and GM1 \cite{1991PhRvL..67.2414G,2000NuPhA.674..553P} (after having adapted these in units used by the code). In figure~\ref{fig01}, we compare and contrast the three selected EOS used in this work in the nuclear and supranuclear regime, relevant for NS cores. Below in section \ref{sec:5}, we shall show how this selection of EOS are not only physically relevant, as it is shown in this section, but also from the astrophysical point of view.

\section{Stability of Equilibrium Models}\label{sec:4}

To solve the system of field equations (\ref{eqA1a}--\ref{eqA1d}) and the hydrostationary equilibrium equation (\ref{eq6}), one has to fix one (in the spherical static case) or two parameters (in the rotating case). The first quantity one has to fix is the central value of energy density, ${\varepsilon}_c$. For a rotating model, one can choose the second parameter amongst different possibilities: the axes ratio ($r_{\rm pol}/r_{\rm eq}$) of coordinate radii, angular velocity ($\Omega$), dimensionless angular momentum ($j$), gravitational mass ($M$) or baryonic mass ($M_b$). Thus, it is always possible to construct a sequence of rotating models by fixing a value of the second parameter and letting the central energy density to vary in a given range which is constrained to stability limits which we now discuss.

The first limit on the stability of uniformly rotating configurations which we take into account is given by the sequence of maximally rotating stars, also referred to as \emph{Keplerian} or \emph{mass-shedding} sequence. In all the stars belonging to such a sequence, the gravitational force equals the centrifugal force at the star equator, in such a way that faster rotation rates would induce the expulsion of mass from the star. The RNS code calculate this sequence by decreasing the axis ratio (which correspond to an increase in angular velocity) until the angular velocity equals the one of a particle orbiting the star at its equator.

Another limit to the physically relevant models is determined by the called secular axisymmetric instability. For static configurations, the maximum stable mass (the critical mass) coincides with the first maximum of a sequence of configurations with increasing central density, namely the first point where $\partial M$/$\partial \varepsilon_c=0$, being $M$ the mass of a configuration with central density $\varepsilon_c$. At this point, the frequency of the radial perturbations vanishes. For higher densities, imaginary frequencies are obtained which lead to non-oscillatory perturbations, hence an instability. Thus, for static configurations, a \emph{turning-point} of the $M$-$\varepsilon_c$ relation locates the onset of unstable configurations. This instability proceeds on \emph{secular} timescales, i.e. not dynamical, so that it proceeds on long times that allow the star to accommodate itself to the energy loss that occurs when going from one equilibrium point to another during gravitational collapse (see, e.g., Ref.~\cite{1988ApJ...325..722F}, and references therein). As shown by Friedman, Ipser and Sorkin in \cite{1988ApJ...325..722F}, the turning-point method leading to points of secular instability can be also used in uniformly rotating stars as follows. In a constant angular momentum sequence, the turning point of a sequence of configurations with increasing central density separates secularly stable from secularly unstable configurations. Namely, secular axisymmetric instability sets in at
\begin{equation}\label{eq:TurningPoint}
\left.\frac{\partial M\left(\varepsilon_c,J\right)}{\partial\varepsilon_c}\right\vert_{J=\rm constant}=0,
\end{equation}
and therefore the curve connecting all the maxima (turning-points) limits the stability region. The intersection of such a limiting curve with the Keplerian sequence gives the fastest possible configuration. It is important to mention that the numerical code adopted (described in the next section) builds sequences of constant dimensionless angular momentum, defined as
\begin{equation}
\label{eq8}
j \equiv \frac{c J}{G M_{\odot}^2},
\end{equation}
which is the quantity we refer to in the sequel. 

The angular momentum $J$ is computed from the definition
\begin{equation}
\label{eq12}
J=\int_{\Sigma} T_{ab} {\phi}^a {\hat{n}}^b dV,
\end{equation}
being $\Sigma$ a spacelike $3$-surface, ${\hat{n}}^a = {\nabla}_a t/\vert {\nabla}_b t {\nabla}^b t \vert$ the unit normal vector field to the $t$=constant spacelike hypersurfaces and $dV = \sqrt{\vert ^{3}g \vert} d^3 x$ the proper 3-volume element (with $^{3}g$ the determinant of the $3$-metric). With this, equation~(\ref{eq12}) becomes \citep{1976ApJ...204..200B}
\begin{equation}
\label{eq13}
J=\int B^2e^{2\zeta - 4\nu} \frac{(\varepsilon + P) v}{1-v^2} r^3 {\sin}^2(\theta) dr d\theta d\phi.
\end{equation}

\section{Mass-Radius Relation, Observational Constraints, and Stability Region}\label{sec:5}

There are in the literature many different numerical schemes and consequently codes to compute relativistic, rotating figures of equilibrium. For the numerical integration of the equilibrium equations we use in this work the public code RNS\footnote{http://www.gravity.phys.uwm.edu/rns/} by Stergiuolas and Friedman \cite{1995ApJ...444..306S}. This code is a numerical implementation based on the scheme by Cook, Shapiro and Teukolsky \cite{1992ApJ...398..203C} (firstly implemented for realistic NS EOS in \citep{1994ApJ...424..823C}), which is a modified version of the method envisaged by Komatsu, Eriguchi and Hachisu \cite{1989MNRAS.237..355K}. We refer the reader to Ref.~\cite{2003LRR.....6....3S} for further details on the numerical schemes.

The major intuitive effect of rotation is to deform the figure of equilibrium with respect to the spherical static counterpart. This can be seen from many points of view. For instance, we can compute sequences of constant angular velocity $\Omega$. An important aspect should be taken into account nowever: the RNS code builds fast rotating models starting from a spherical (static) guess and decreases the polar to equatorial radii ratio until the parameter fixed (e.g.~the angular velocity) is reached with a prescribed accuracy. Thus, the axes ratio is a parameter used intrinsically by the numerical method, while other parameters (see the beginning of previous section for a list) can be chosen, but are reached spanning decreasing values of axis ratio. In particular, as an example, the code does not converge for every value of fixed angular velocity in every range of central energy density, and the range of convergence gets reduced by decreasing the angular velocity. To be more precise, choosing fixed rotation frequencies below 300~Hz, the code fails to converge in the entire range in which equilibrium models should exist (thus between the Keplerian and secular instability limits), even adopting a very dense numerical grid (300 angular times 600 radial points), different accuracy and tolerance values (down to $10^{-16}$) or values of a relaxation factor from 1 to 0.8. Effectively the code does converge for this kind of rotation frequencies, but in very limited ranges of energy density. Thus, how can the slow rotation regime be recovered in all the stability region? As a technical advise, we would like to mention that one could compute first, sequences keeping various values of axes ratio constant (in the vicinity of unity), and then select in this set of models the ones with small values of a particular angular velocity. We construct with this simple method the sequences with low rotation frequency (e.g.~models from 50~Hz to 200~Hz). For other values of rotation frequency and for all other parameters constant sequences, we achieved optimal convergence using a 300 angular times 600 radial points numerical grid, and accuracy and tolerance of $10^{-8}$, while the relaxation parameter was not necessary.

\begin{figure*}[!hbtp]
\includegraphics[width=0.32\hsize,clip]{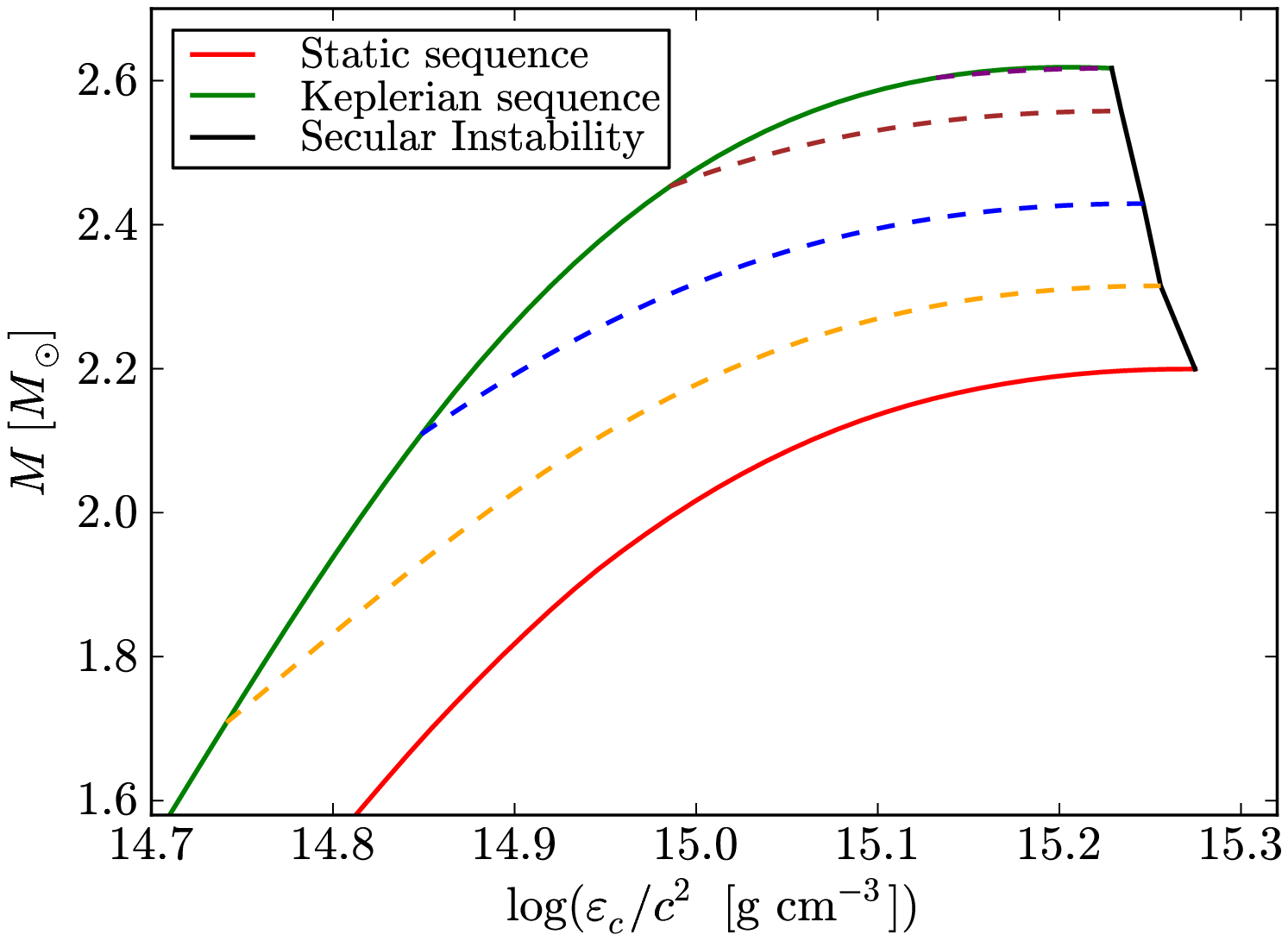}
\includegraphics[width=0.32\hsize,clip]{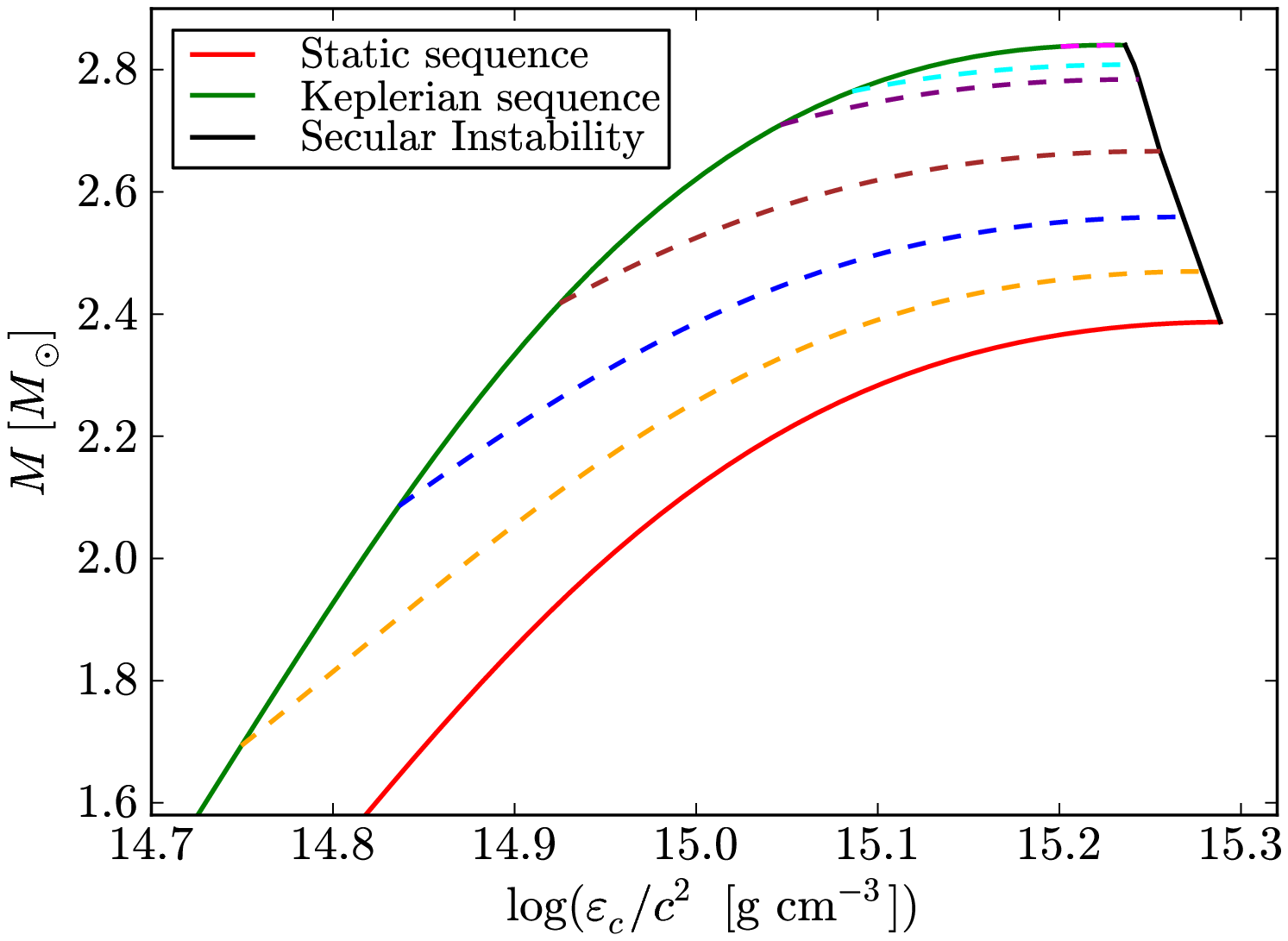}
\includegraphics[width=0.32\hsize,clip]{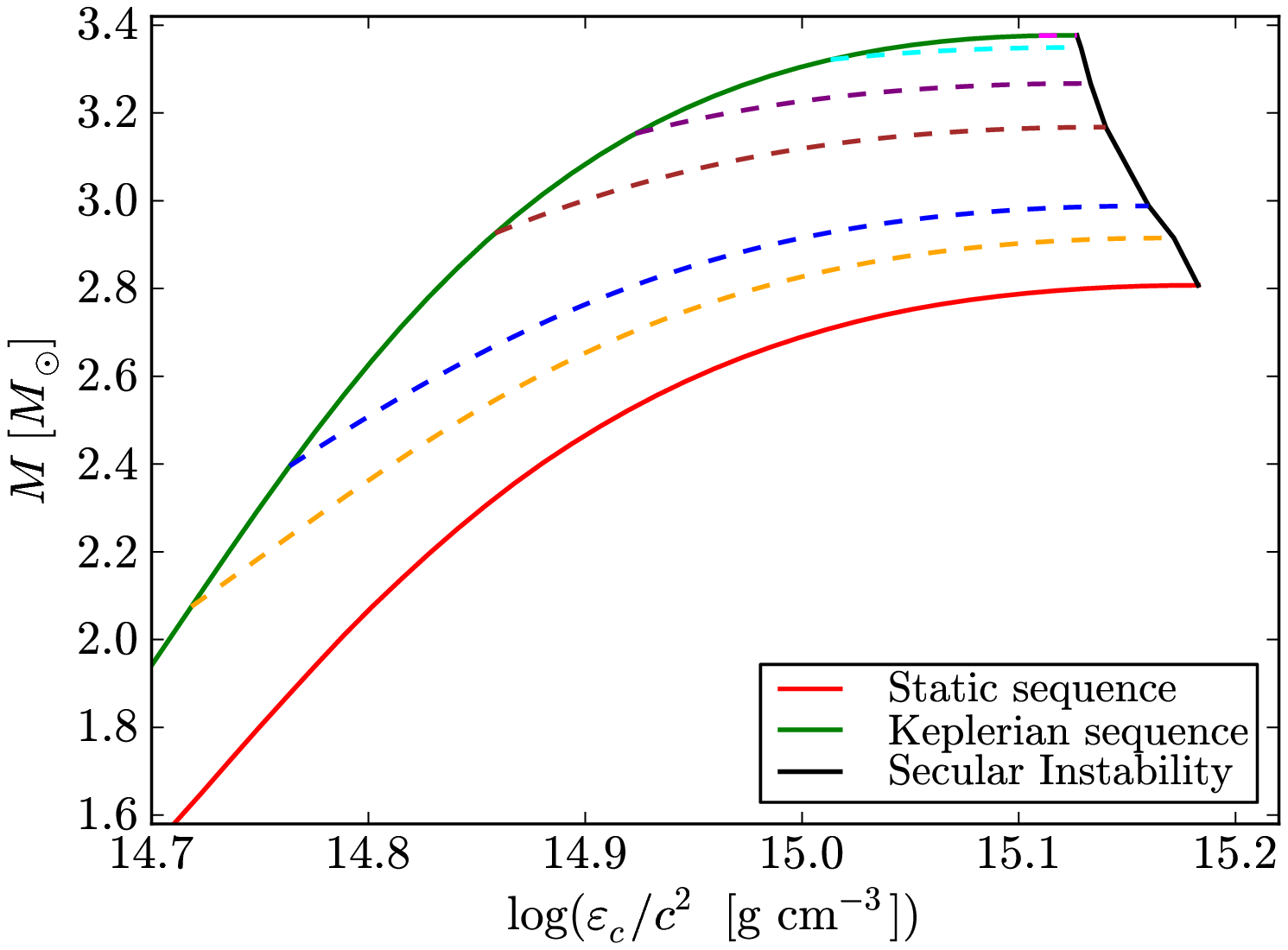}
\caption{\label{fig02}Gravitational mass is plotted against central energy density for $j$-constant sequences obtained with the EOS TM1, GM1 and NL3 (from top to bottom). In this plot and hereafter, the red, green and black curves represent respectively the static sequence, the Keplerian sequence, and the limit for secular stability. Here other colors stand for various $j$-constant sequences.}
\end{figure*}

Figure~\ref{fig02} shows the total mass-central energy density plane for the selected EOS TM1, GM1 and NL3. We also show the stability limits discussed above in section \ref{sec:4} and show explicitly some constant angular momentum sequences.

\begin{figure}[!hbtp]
\includegraphics[width=0.8\hsize,clip]{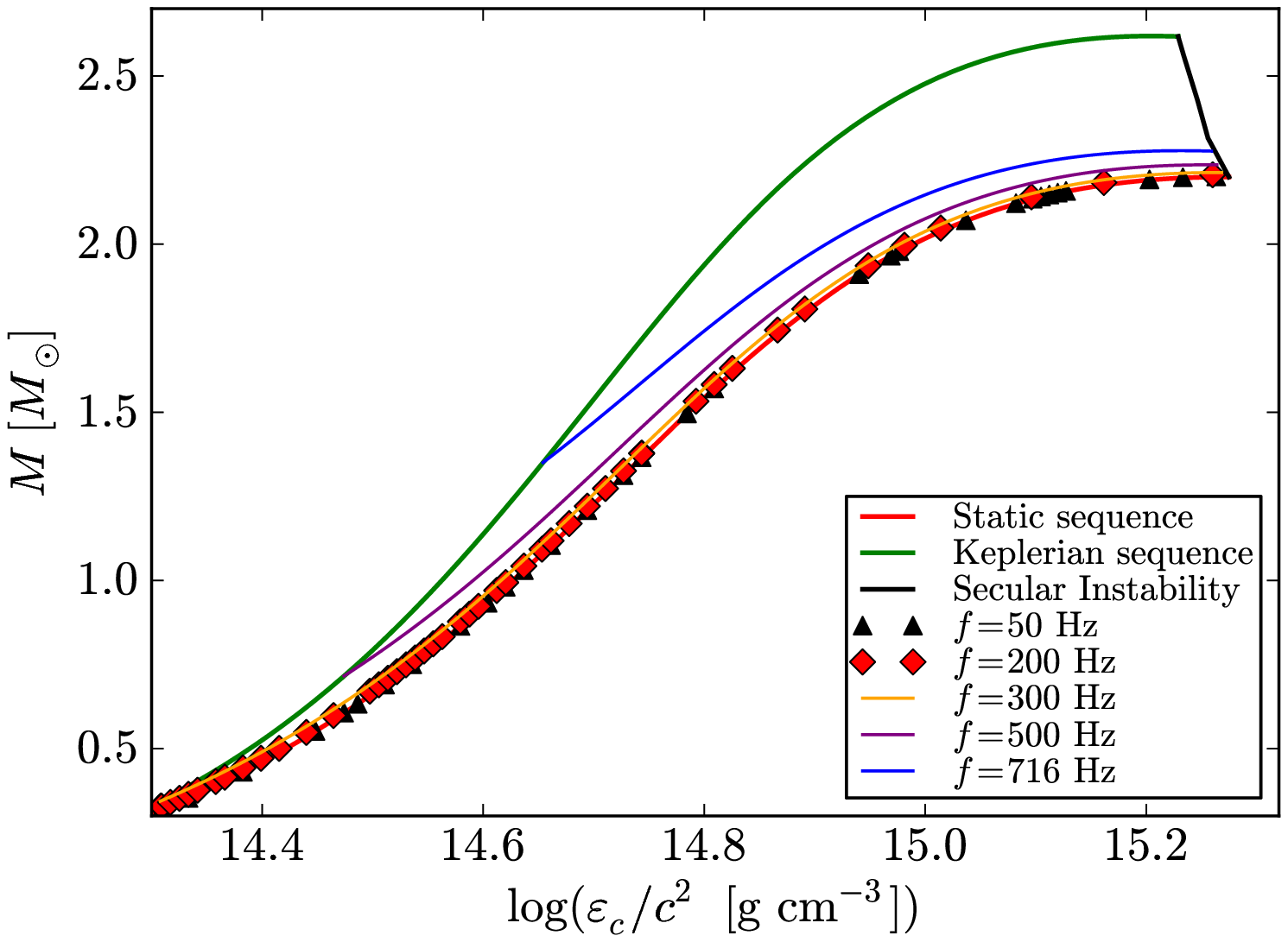}
\includegraphics[width=0.8\hsize,clip]{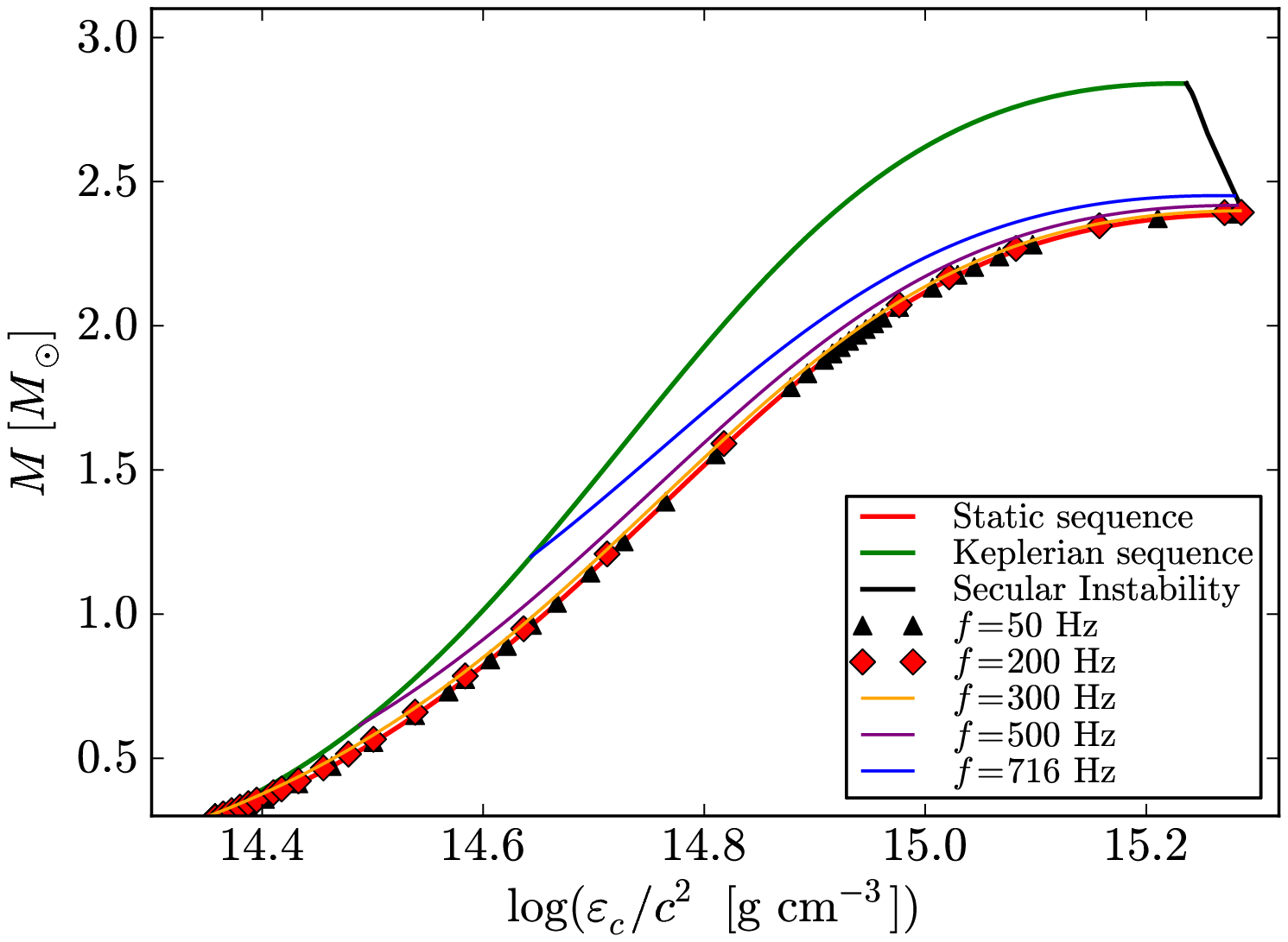}
\includegraphics[width=0.8\hsize,clip]{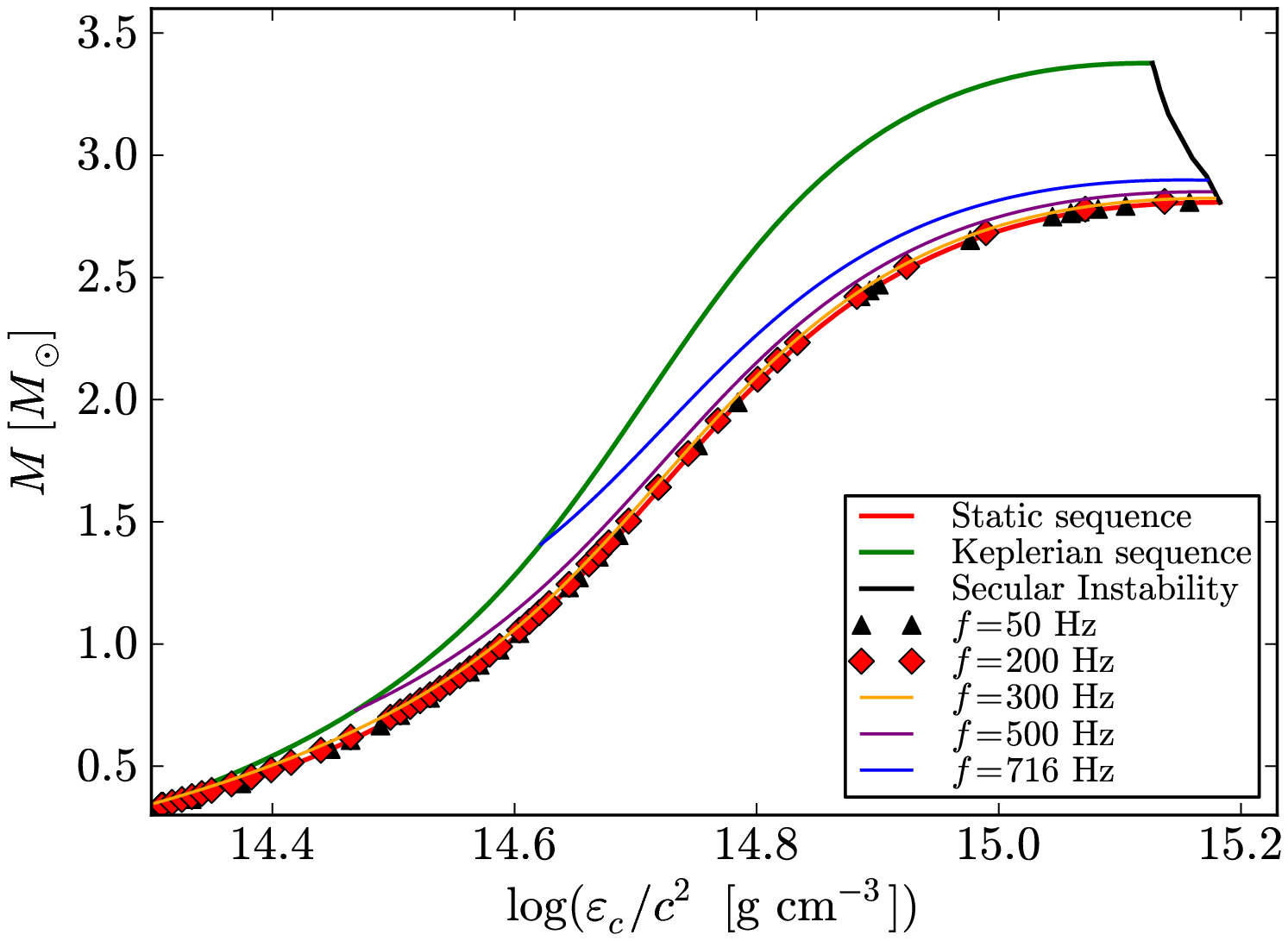}
\caption{\label{fig03}Mass versus central energy density using the EOS TM1, GM1 and NL3 (from top to bottom) set of parameters. Red and green curves represent the static and Keplerian sequences. Other colors correspond to constant frequency sequences of value $716$~Hz (fastest observed pulsar; blue), $500$~Hz (purple), $300$~Hz (orange), $200$~Hz (red diamonds) and $50$~Hz (black triangles).}
\end{figure}

Figure~\ref{fig03} shows instead the total mass-central energy density plane but in this case we show explicitly some selected constant rotation frequency sequences ranging from 50~Hz all the way up to the rotation frequency of the fastest observed pulsar, PSR J$1748-2446$ad, with $f=\Omega/(2\pi)\approx 716$~Hz \cite{2006Sci...311.1901H}.

\begin{figure}[!hbtp]
\includegraphics[width=0.8\hsize,clip]{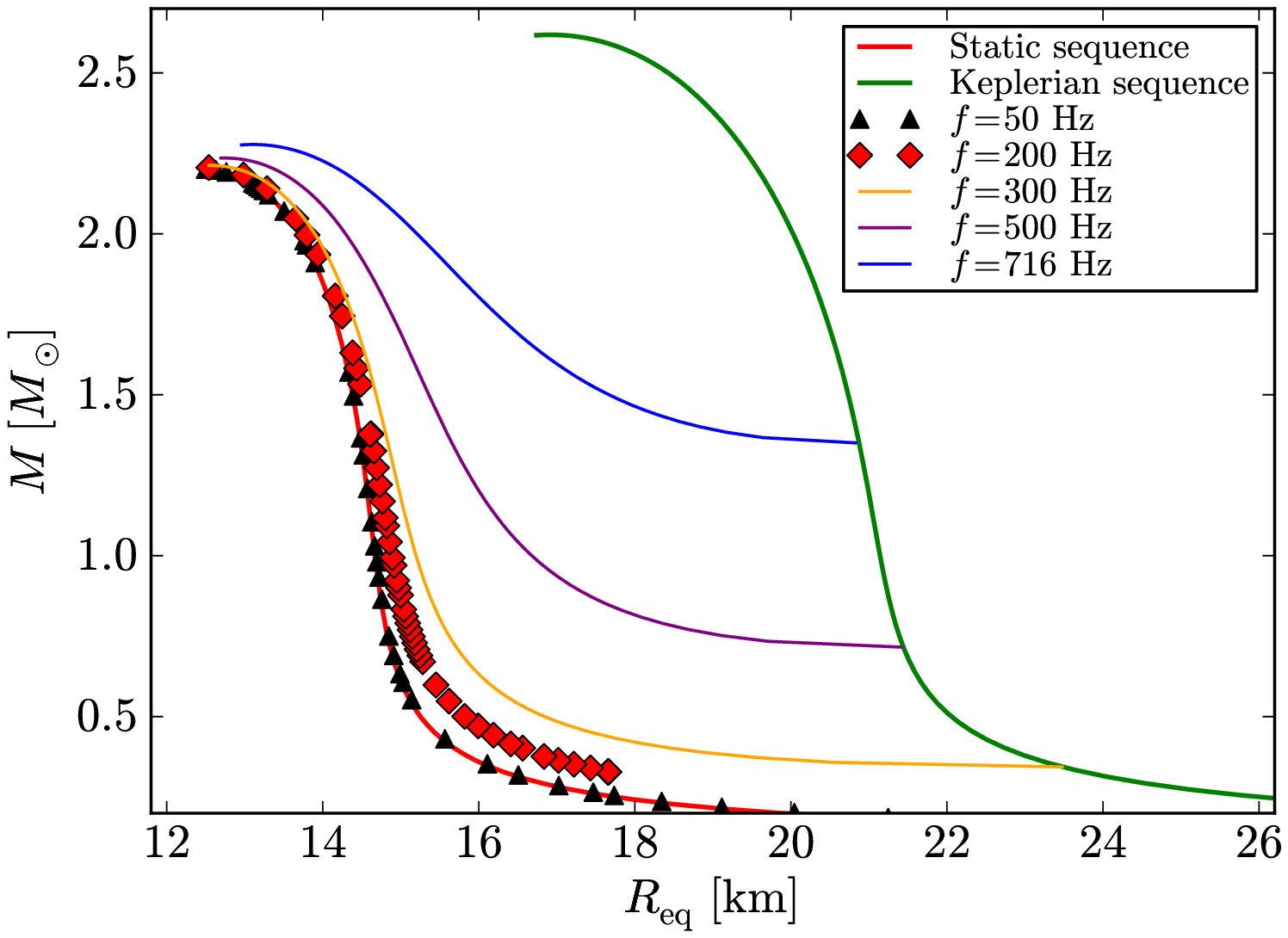}
\includegraphics[width=0.8\hsize,clip]{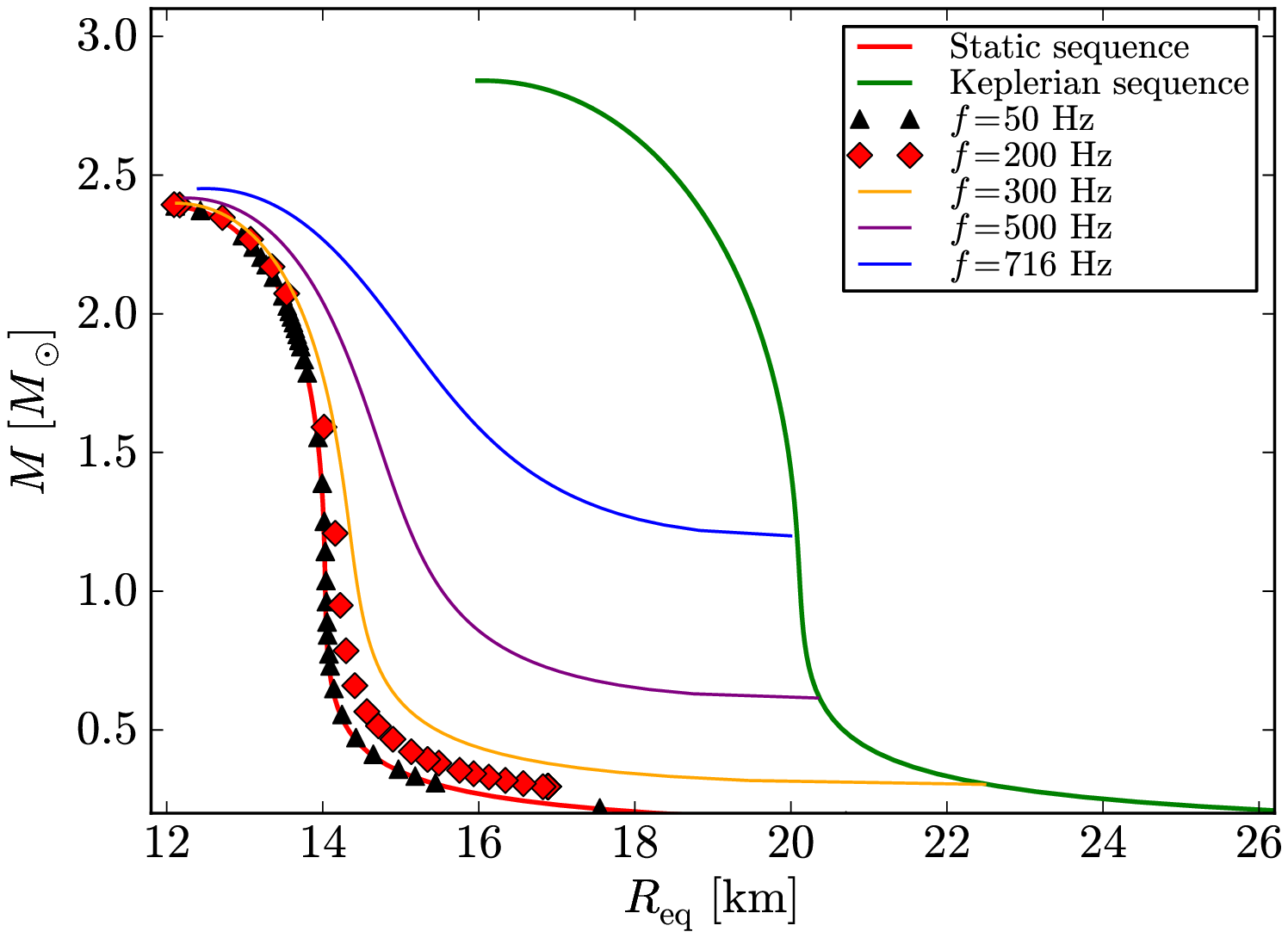}
\includegraphics[width=0.8\hsize,clip]{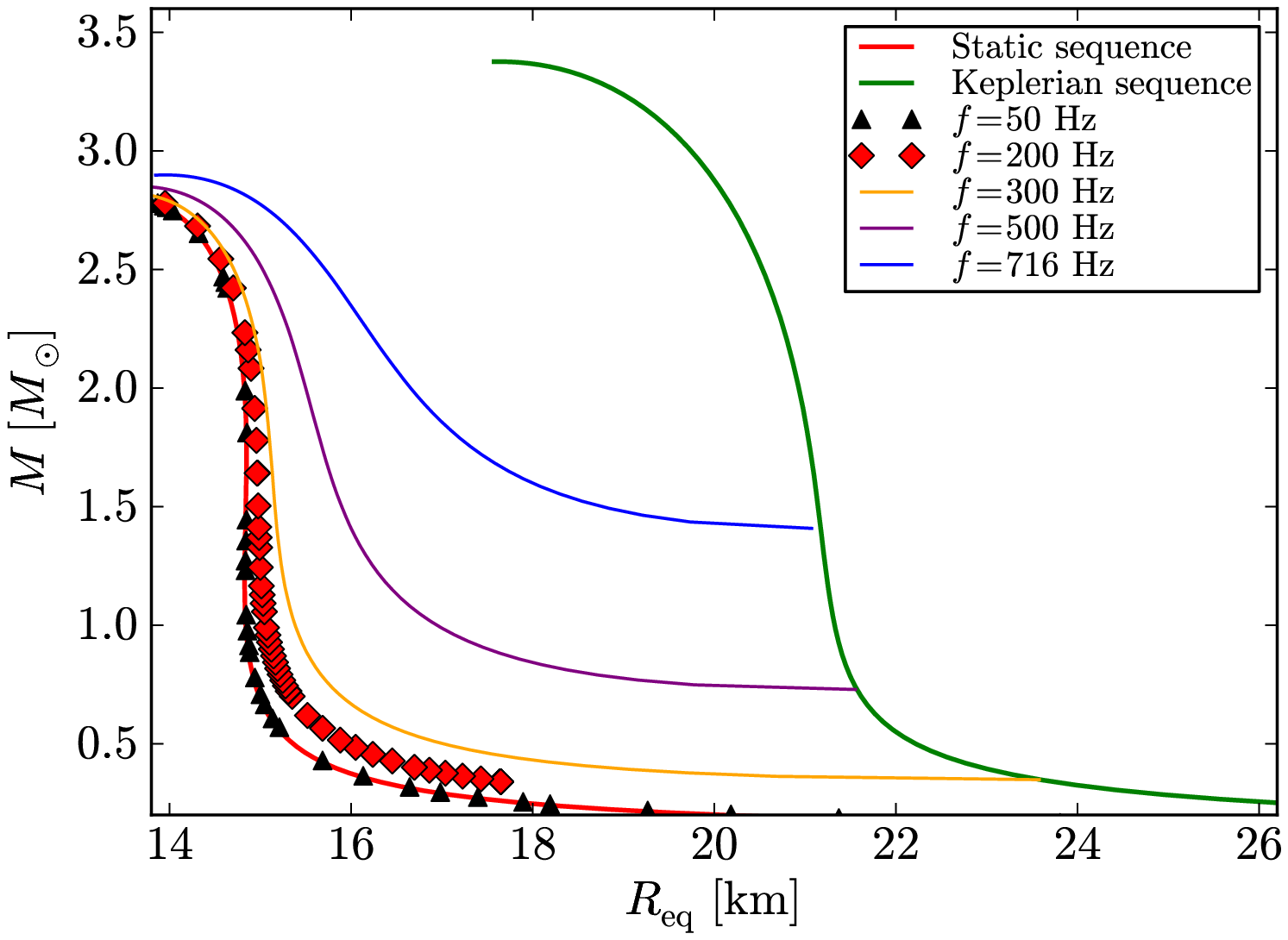}
\caption{\label{fig04}Mass versus equatorial radius using EOS TM1, GM1 and NL3 (from top to bottom) set of parameters for same sequences as in figure~(\ref{fig03}). The convention of the plot colors and symbols is the same as in figure~\ref{fig03}.}
\end{figure}

In figure~\ref{fig04} we plot the same $\Omega-$constant sequences to show the relation between $M$ and the equatorial radius, $R_{\rm eq}$.

\begin{figure}[!hbtp]
\includegraphics[width=\hsize,clip]{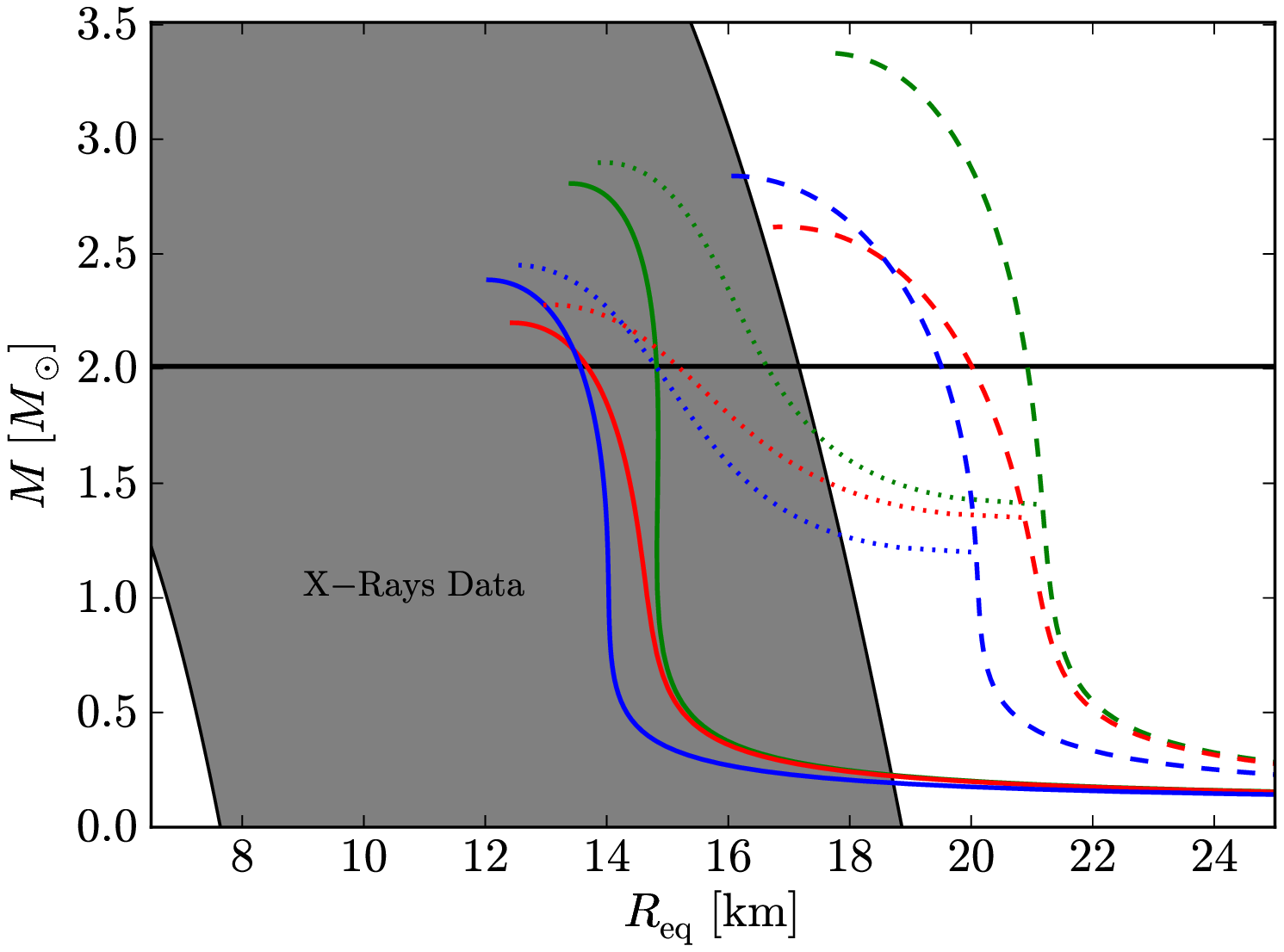}
\caption{Mass versus equatorial radius relation using the TM1, GM1 and NL3 EOS (respectively with colors red, blue and green), plotted together with up-to-date observational constraints. Solid color curves represent static NS configurations; dotted color curves represent the sequence of models rotating with spin frequency of the fastest observed pulsar (PSR J1748--2446ad), $f=716~Hz$, and dashed color curves represent sequences of models rotating at the Keplerian frequency. The gray-shaded region corresponds to the constraints given by the X-ray data while the horizontal lines are the lower and upper bounds to the mass of the most massive observed pulsar, PSR J1614--2230, namely $M=2.01 \pm 0.04 M_\odot$. Further details of these constraints can be found in the text.}\label{figOBS}
\end{figure}

With the knowledge of the mass-radius relations predicted by the theory, we are now in the position of comparing and contrasting them with existing observational constraints, in order to validate the selection of EOS of the present work, not only from the already presented physical aspects, but also from the astrophysical ones. Current observational constraints to the mass-radius relation of NSs are (see figure~\ref{figOBS}):

\begin{itemize}

\item 

\emph{Most massive NS observed}. The mass value of the most massive NS observed is the one of PSR J1614--2230 with $2.01 \pm 0.04 M_\odot$ \cite{2010Natur.467.1081D}. The rotation frequency of this pulsar is 46~Hz; thus the deviations from spherical symmetric are negligible. This implies that every mass-radius relation for non-rotating NSs must have a maximum stable mass larger than this value.

\item 

\emph{Fastest observed NS}. The highest rotation frequency observed in a pulsar is the one of PSR J$1748-2446$ad with $f=716$~Hz \cite{2006Sci...311.1901H}. The constant frequency sequence of this value for any mass-radius relation must have at least one stable configuration that supports such a rotation frequency; namely the constant frequency sequence for this pulsar must lie within the region of stability. This is actually a very weak constraint since most of NS models allow much higher rotation frequencies. Interestingly, as we shall show below, the construction of the constant frequency sequence for PSR J$1748-2446$ad allows us to infer a lower mass for this pulsar.

\item 

\emph{Constraints to the NS radius}. Since the surface temperatures of not-so-young NSs ($t>10^3$--$10^4$~y) are of the order of million degrees (see, e.g.,~\cite{2014PhRvC..90e5804D}), their thermal spectrum is expected to peak in the soft X-rays. Thus, the modeling of the NS X-ray emission appears to be, at the present, one of the most promising methods to obtain information on the NS radius. Systems which are currently used to this aim are isolated NSs, quiescent low-mass X-ray binaries (qLMXBs), NS bursters, and rotation-powered millisecond pulsars (see, Ref.~\cite{2015A&A...576A..68F}, and references therein). From the modeling of the observed spectrum, the radius of the NS as measured by an observer at infinity, $R_{\infty}=R/\sqrt{1-2 G M/(c^2 R)}$, can be extracted\footnote{Actually, accurate spectra modeling leads to preferable values for both mass and radius; however, for a simpler comparison between different results from different methods and for a simple test of the mass-radius relation it is sufficient to plot the constraints obtained from the values of $R_{\infty}$ consistent with the data \cite{2006ApJ...644.1090H}.}. The observation of a preferable radius at infinity clearly represents a constraint to the NS mass-radius relation since the above definition for $R_{\infty}$ can be rewritten as $2 G M/c^2 = R-R^3/(R_{\infty}^3)$. In Ref.~\cite{2014EPJA...50...40L} (see, also, Ref.~\cite{2014MNRAS.444..443H}), the X-ray emission from the NSs in the qLMXBs M87, NGC 6397, M13, $\omega$ Cen, and NGC 6304 was revisited, and in Ref.~\cite{2006ApJ...644.1090H} the one of the NS X7 in the Globular Cluster 47 Tucanae. From the extracted values of $R_{\infty}$ consistent with these observational data at 90\% confidence level, we can conclude that the current X-ray data constraints very weakly the mass-radius relation, allowing radii in the interval $R_{\infty}=[7.64,18.86]$, where the lower limit is obtained for NGC 6304 and the upper one for X7. It is important to mention that X-ray measurements suffer from a variety of uncertainties which are the main reason for the very large spread in possible NS radii. The spectra modeling depends on the atmosphere composition, magnetic fields, accurate knowledge of the distance to the source, hence the extinction, and to some extent on the NS exterior geometry which could be affected by the rotation of the NS in the case of some LMXBs which could harbor NSs rotating with frequencies of a few hundreds of Hz (see, e.g., Ref.~\cite{2015ApJ...799...22B}, for details). In these latter cases, a more reliable comparison between theory and the above data constraints, which assume spherical symetry, could be obtained by ploting mass-radius relation using, instead of the equatorial radius, a mean or average spherical radius such as the authalic radius, $\langle R \rangle = (2 R_{\rm eq} + R_{\rm pol})/3$. However, for the purposes of this work, it is sufficient to make a comparison with the mass-radius relation produced by the non-rotating configurations.

\end{itemize}

An additional constraint to the mass-radius relation might come from the request of causality to the EOS, namely the condition that the speed of sound in the NS interior cannot exceed the speed of light. However, for the present set of EOS this condition is automatically satisfied by construction since the models are relativistic. One can therefore see from figure~\ref{figOBS} that the spherical (static) models (solid curves) obtained by the EOS selection of this work are well in agreement with the current constraints of the NS mass-radius relation determined by most updated observational data.

An often useful physical quantity to be computed is the binding energy of the configurations, or the relation between the baryonic mass and the gravitational mass. For non-rotating NSs, we found that for the three analyzed EOS, the following relations hold:
\begin{eqnarray}\label{eq:MbM}
\frac{M_b}{M_\odot} &\approx& \frac{M}{M_\odot}+\frac{13}{200}\left(\frac{M}{M_\odot}\right)^2,\nonumber \\ \frac{M}{M_\odot}&\approx& \frac{M_b}{M_\odot}-\frac{1}{20}\left(\frac{M_b}{M_\odot}\right)^2,
\end{eqnarray}
where $M_b$ is the baryonic mass, hold, and thus appear to be a universal property. The maximum relative errors obtained for non-rotating sequences of GM1, TM1 and NL3 are respectively $1.4\%$, $1.3\%$ and $0.99\%$. For rotating configurations, $M=M(M_b,J)$ or $M_b=M_b(M,J)$, we find that for the our set of EOS there is indeed a common relation given by
\begin{equation}
\frac{M_b}{M_\odot}=\frac{M}{M_\odot}+\frac{13}{200}\left(\frac{M}{M_\odot}\right)^2\left(1 - \frac{1}{130} j^{1.7}\right),
\end{equation}
accurate within an error of 2\%, and which duly generalizes equation~(\ref{eq:MbM}).

Turning back to the above plots, we can clearly see that, as expected, the higher the frequency of rotation is, the higher results the value of the mass at which start the departures from the non-rotating mass-radius relation. We find that for rotation frequencies $\lesssim 200$~Hz (or rotation periods $\gtrsim 5$~ms), the non-rotating star becomes an accurate representation of the object. This is in accordance with previous results, see e.g. figure 5 in Ref.~\cite{benhar05}, where it was shown that the moment of inertia of sequences computed with different EOS, start to deviate considerably from the static and the slow-rotation Hartle's approximations for frequencies above $\sim 0.2$~kHz. As we show below, this is also the case for the moment of inertia in the same range of frequencies (thus, the moment of inertia of non-rotating configurations, can be safely approximated with the one of spinning configurations, with frequencies below the aforementioned limit, and viceversa). For higher frequencies, full rotation effects are needed for an accurate description. This is especially important for objects with masses lower that the maximum value, where departures from a non-rotating or slow rotation approximation become more and more evident. 

Following this reasoning, it is important to see how a constant frequency sequence imposes structure constraints to a pulsar. Particularly interesting becomes the case of the $f=716$~Hz sequence (blue curve), which corresponds to the fastest observed pulsar, PSR J$1748-2446$ad. The constant frequency sequence intersects the stability region in two points: at the maximally rotating Keplerian sequence, defining a minimum mass for the pulsar, and at the secular axisymmetric instability limit, in the upper part, defining the maximum possible mass for the given frequency. Clearly these minimum and maximum mass values depend upon the EOS. For the EOS employed here, we can see that the mass of PSR J$1748-2446$ad has to be $\gtrsim [1.41,1.35,1.20]~M_\odot$ for NL3, TM1, and GM1, respectively.

\begin{figure}[!hbtp]
\includegraphics[width=0.8\hsize,clip]{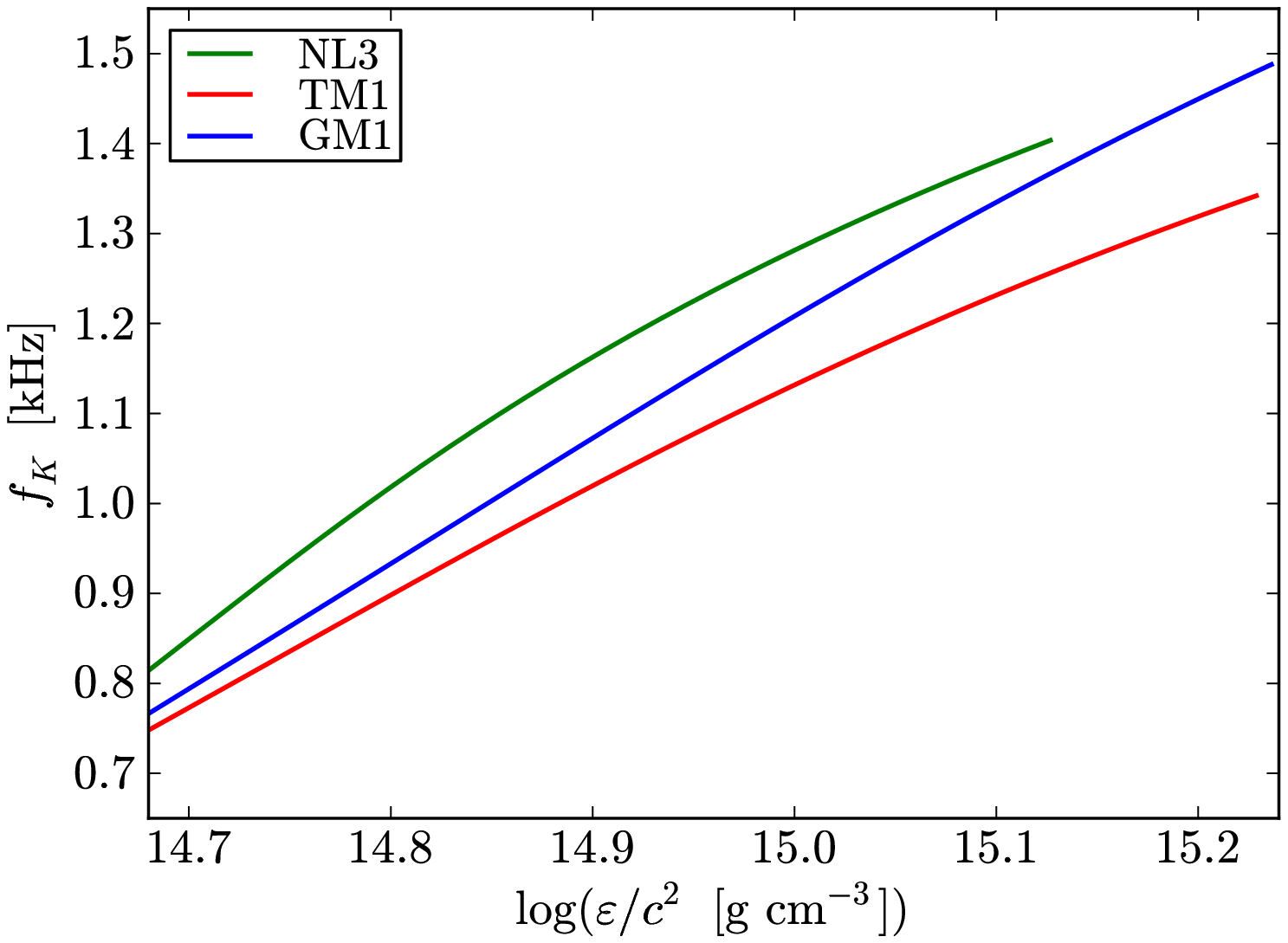}
\caption{\label{fmax}Frequency of the maximally rotating configurations (Keplerian sequence) as a function of the total NS mass for the TM1, GM1 and NL3 EOS. The curves end at the maximum frequency configuration, which is located at the intersection between the Keplerian and the secular axisymmetric instability sequences.}
\end{figure}

We now determine the maximum rotation frequency of NSs. The fastest configuration for a given EOS is the one that terminates the Keplerian sequence, namely the configuration at the intersection between the Keplerian and the secular axisymmetric instability sequences. We show in figure~\ref{fmax} the rotation frequency of the maximally rotating configurations, i.e. the frequencies of the NSs along the Keplerian sequence.

\begin{figure}[!hbtp]
\includegraphics[width=\hsize,clip]{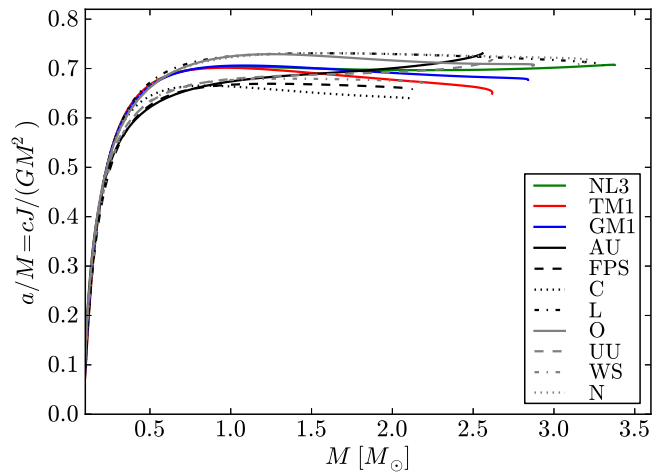}
\caption{\label{aoverM}Dimensionless angular momentum (``Kerr parameter''), $a/M\equiv c J/(G M^2)$, as a function of the total NS mass along the Keplerian sequence for the EOS selected in this work (colored curves). For comparison we show the results for additional EOS, taken from a set supplied by the RNS code (EOS.INDEX file). We refer the reader to the RNS web page and references therein for further details of these EOS.}
\end{figure}

Another important quantity for this discussion is the dimensionless angular momentum (``Kerr parameter''), $a/M\equiv c J/(G M^2)$, which we show in figure~\ref{aoverM} as a function of the total mass for the maximally rotating configurations, namely the Keplerian sequence. It can be seen how the maximum value attained by the NS, $(a/M)_{\rm max}\approx 0.7$, holds for all the selected EOS. The maximum value is reached for the mass $[0.96,1.05,3.37]~M_\odot$ for the TM1, GM1, and NL3 EOS, respectively. The existence of such a particular maximum, EOS-independent, value of $a$ possibly implies the existence of universal limiting values of the NS compactness and the rotational to gravitational energy ratio. This is a conjecture which deserves further exploration. In the same plot, the same sequences obtained with other already known EOS (represented by differently dashed curves), obtained assuming widely different kind of interactions and via different many-body-theories, are shown, and the reader can notice a general universal behavior of the dimensionless angular momentum, even if for these other EOS the exact maximal values of this dimensionless parameter are slightly different. On the other hand, such a general behavior of $a$ parameter in not surprising, in fact, as it was already shown in Ref.~\cite{2014PhRvL.112t1102C}, it can be chosen as a parameter to establish a universal I-Love-Q relation. Nevertheless the important argument here is that, although different stiffness, our chosen set of EOS presents a common maximal dimensionless angular momentum $a/M$.

In Table \ref{tab:propNS} we summarize a few relevant quantities of NSs; the maximum stable mass in the non-rotating case, the maximum mass in the case of uniform rotation, the maximum rotation frequency, and the maximum value of the dimensionless angular momentum.

\begin{table}[!hbtp]
\caption{\label{tab:propNS}Some properties of NSs for the selected EOS: critical mass for non-rotating case, $M^{J=0}_{\rm max}$, maximum mass in uniform rotation, $M^{J \neq 0}_{\rm max}$, maximum rotation frequency, $f_{\rm max}$, and maximum dimensionless angular momentum (``Kerr parameter''), $(a/M)_{\rm max} \equiv [c J/(G M^2)]_{\rm max}$.}
\begin{ruledtabular}
\begin{tabular}{ccccc}
EOS & $M^{J=0}_{\rm max}~[M_\odot]$ & $M^{J \neq 0}_{\rm max}~[M_\odot]$ & $f_{\rm max}$~[kHz] & $(a/M)_{\rm max}$\\
TM1 & 2.20 & 2.62 & 1.34 & 0.70 \\
GM1 & 2.39 & 2.84 & 1.49 & 0.71 \\
NL3 & 2.81 & 3.38 & 1.40 & 0.71
\end{tabular}
\end{ruledtabular}
\end{table}

Before closing this section, we would like to provide a formula, useful for astrophysical applications, for the masses of the NSs lying along the secular axisymmetric instability line. Using the dimensionless angular momentum $j$, defined in equation (\ref{eq8}) and related to the Kerr parameter by $j= (M/M_\odot)^2\,a$, we obtained
\begin{equation}\label{eq:secformula}
M=M^{J=0}_{\rm max}(1+k j^l),
\end{equation}
where the values of $M^{J=0}_{\rm max}$ are given in Table \ref{tab:propNS}, $k=[0.017,0.011,0.0060]$ and $l=[1.61,1.69,1.68]$ for the EOS TM1, GM1, NL3, respectively. The maximum relative errors obtained for values of mass along the secular axisymmetric instability line with respect to fits for each EOS are respectively $[0.33\%,0.44\%,0.45\%]$.

\section{Eccentricity and Moment of Inertia}\label{sec:6}

In order to see how a figure of equilibrium becomes deformed by rapid rotation, we compute the eccentricity
\begin{equation}
\label{eq10}
\epsilon = \sqrt{1 - \left( \frac{R_{\rm pol}}{R_{\rm eq}} \right)^2},
\end{equation}
which we plot as a function of the mass, $M$, for the same constant $\Omega$ sequences of the previous figures.

\begin{figure*}[!hbtp]
\includegraphics[width=0.32\hsize,clip]{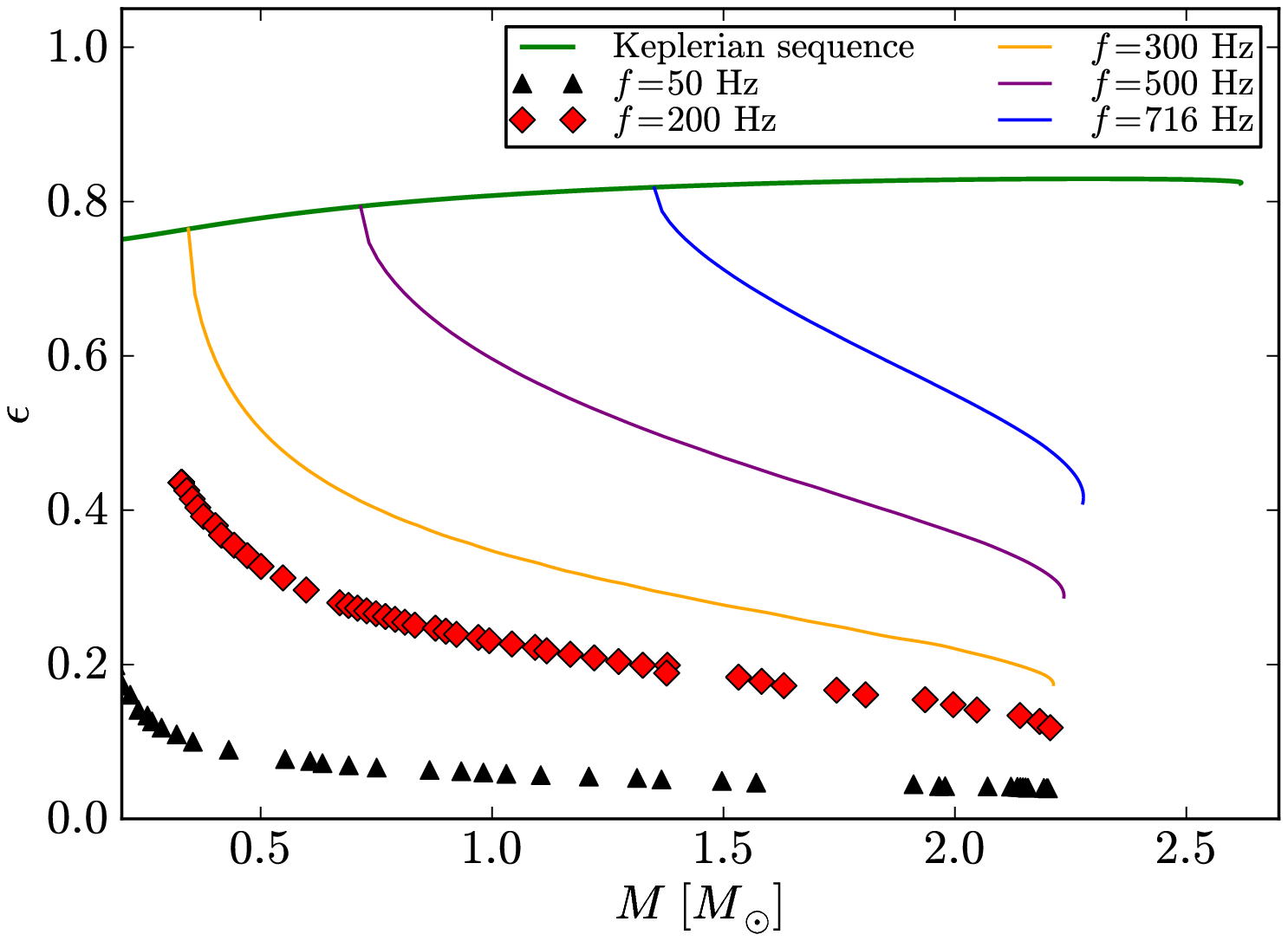}
\includegraphics[width=0.32\hsize,clip]{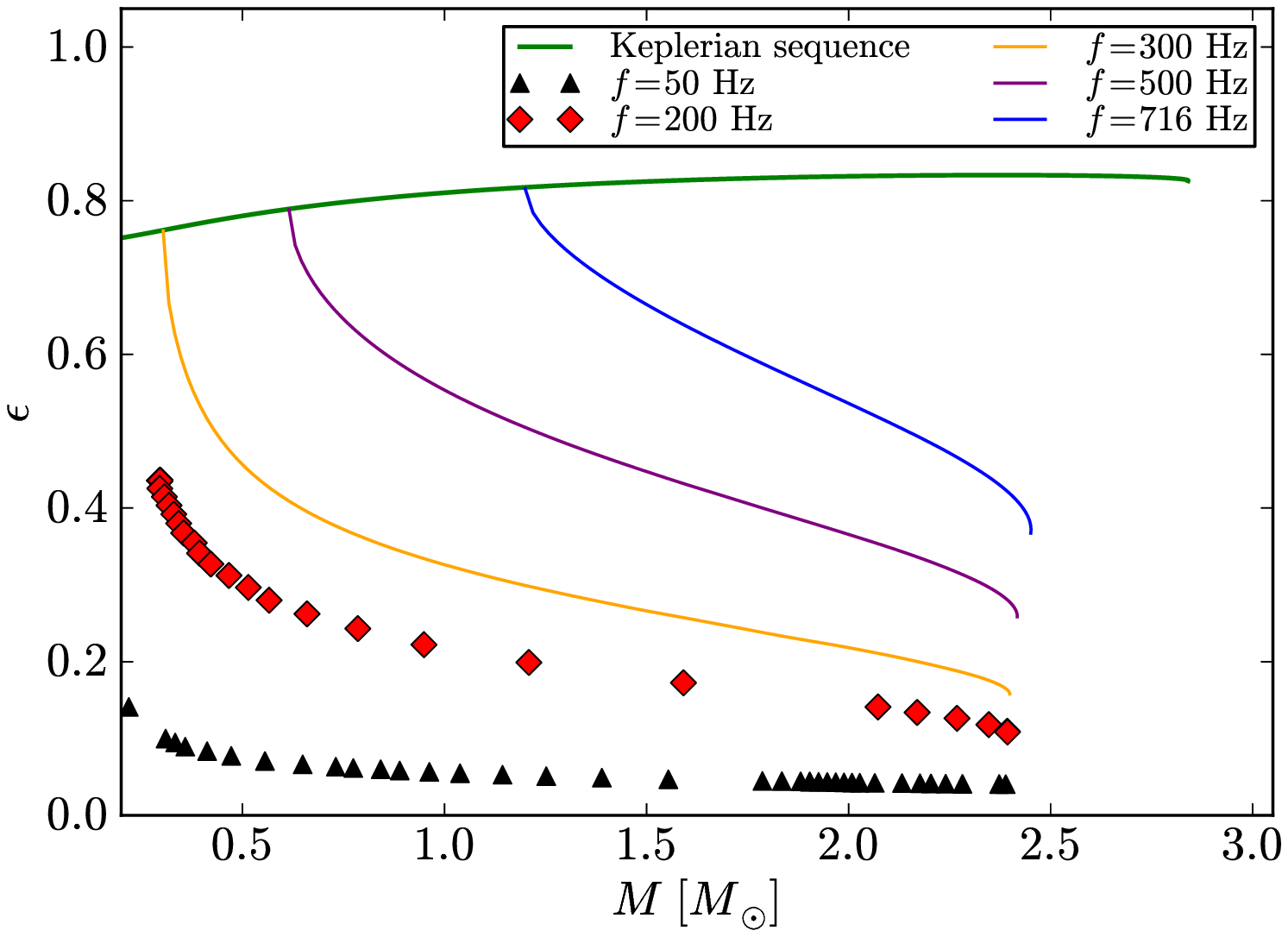}
\includegraphics[width=0.32\hsize,clip]{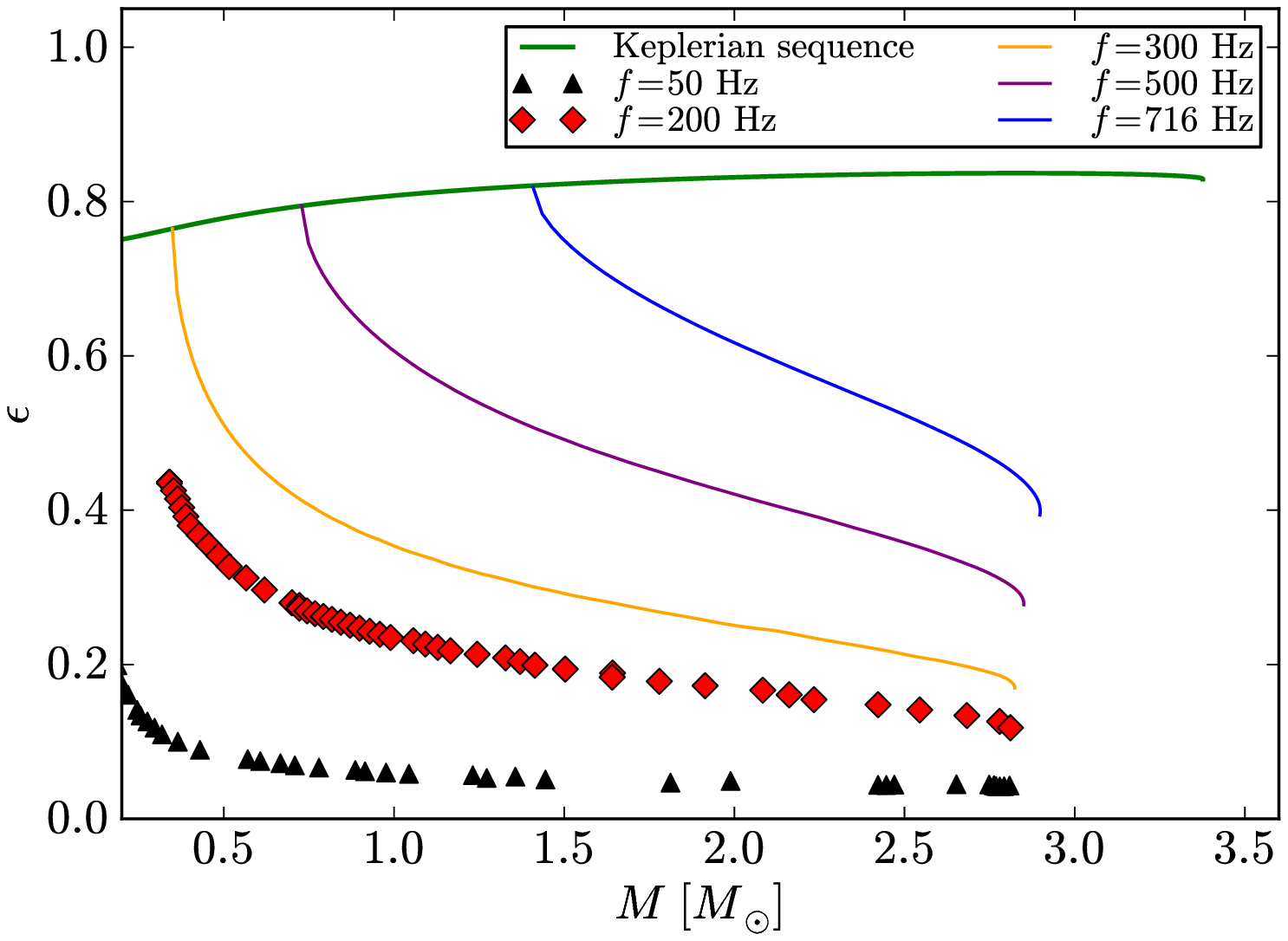}
\caption{\label{fig06}Eccentricity versus gravitational mass using EOS with TM1, GM1 and NL3 (from left to right) set of parameters for same sequences as in figure~\ref{fig03}.}
\end{figure*}

\begin{figure*}[!hbtp]
\includegraphics[width=0.4\hsize,clip]{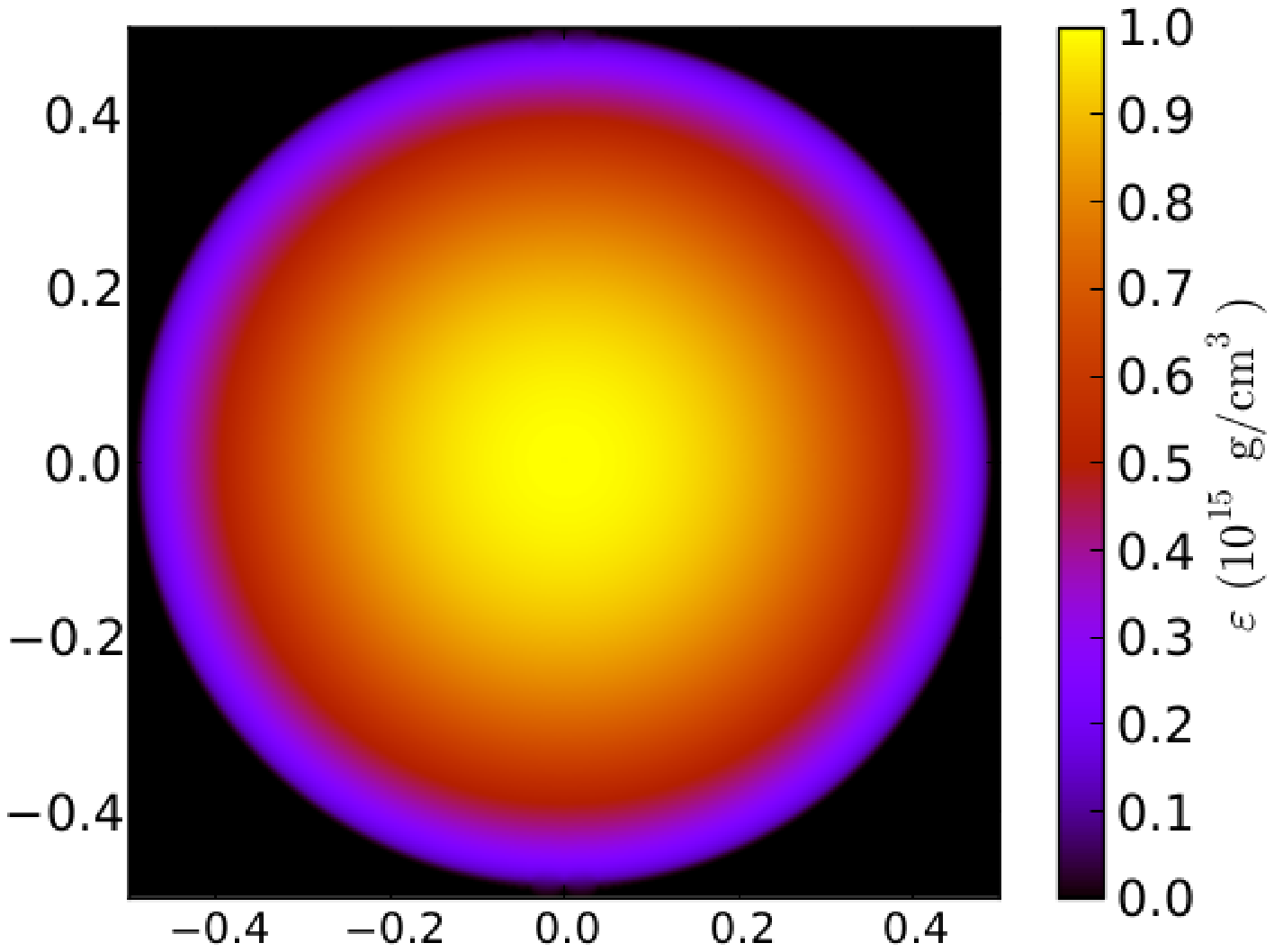}
\includegraphics[width=0.4\hsize,clip]{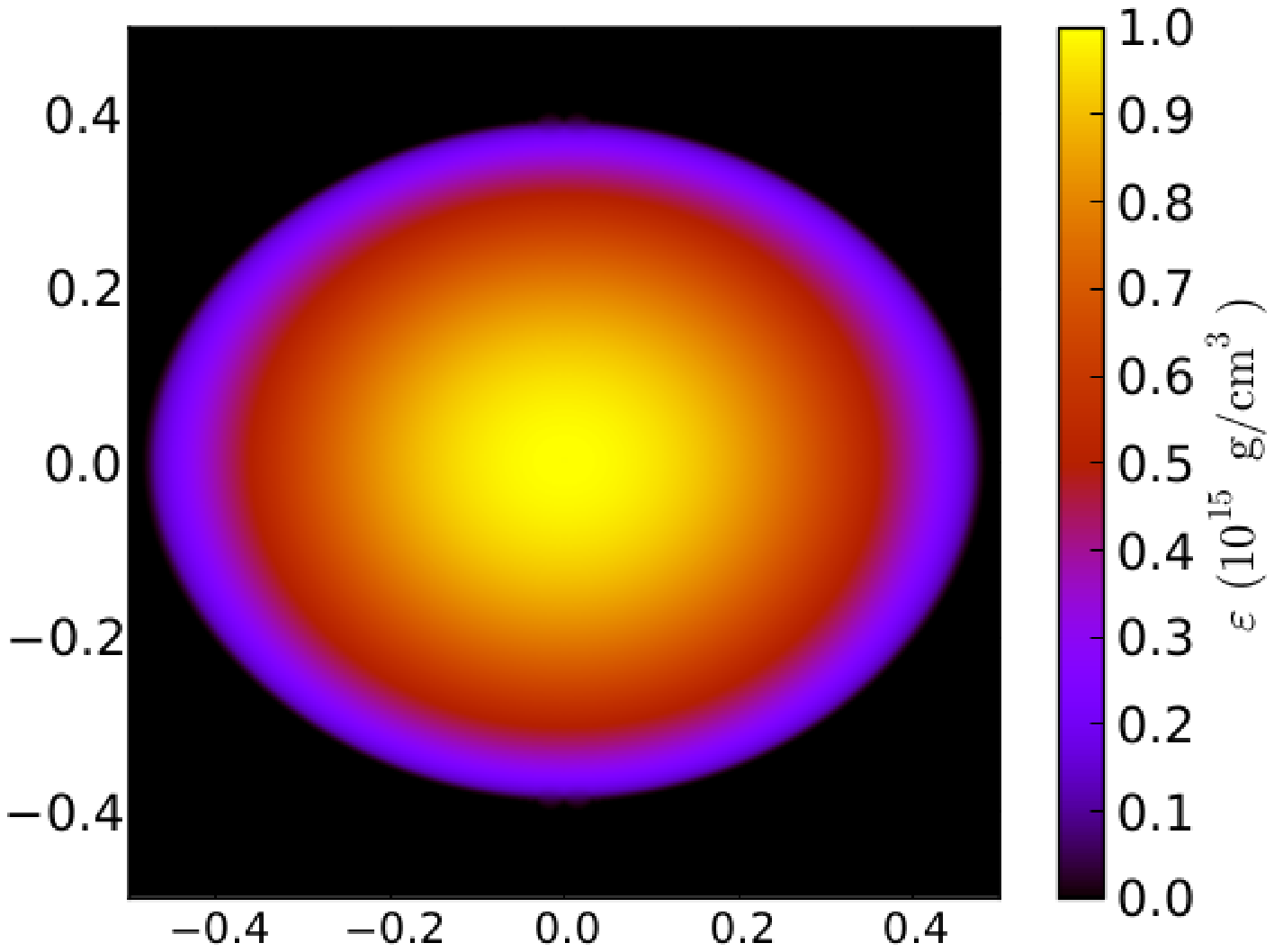}
\caption{\label{fig08}Contours of constant energy density of a model with central value $\varepsilon_c = 10^{15}$~g~cm$^{-3}$ both in the static case (left plot) and in the rotational one with dimensionless angular momentum $j=4$ (right plot) for the GM1 EOS.}
\end{figure*}

It is also interesting to investigate the distribution of the energy density within the figure of equilibrium both the static and rotational case for the different EOS. In figure~\ref{fig08}, we show the contours of constant energy density of a model with central value $\varepsilon_c = 10^{15}$~g~cm$^{-3}$ both in the static case and in the rotational one with dimensionless angular momentum $j=4$, for the sake of example for the GM1 EOS. 

We turn now to compute the moment of inertia of the star, which is one of the most relevant properties in pulsar analysis. The moment of inertia can be estimated as \cite{2003LRR.....6....3S}
\begin{equation}
\label{eq11}
I=\frac{J}{\Omega},
\end{equation}
where $J$ is the star angular momentum which is given by equation~(\ref{eq13}).

\begin{figure*}[!hbtp]
\includegraphics[width=0.32\hsize,clip]{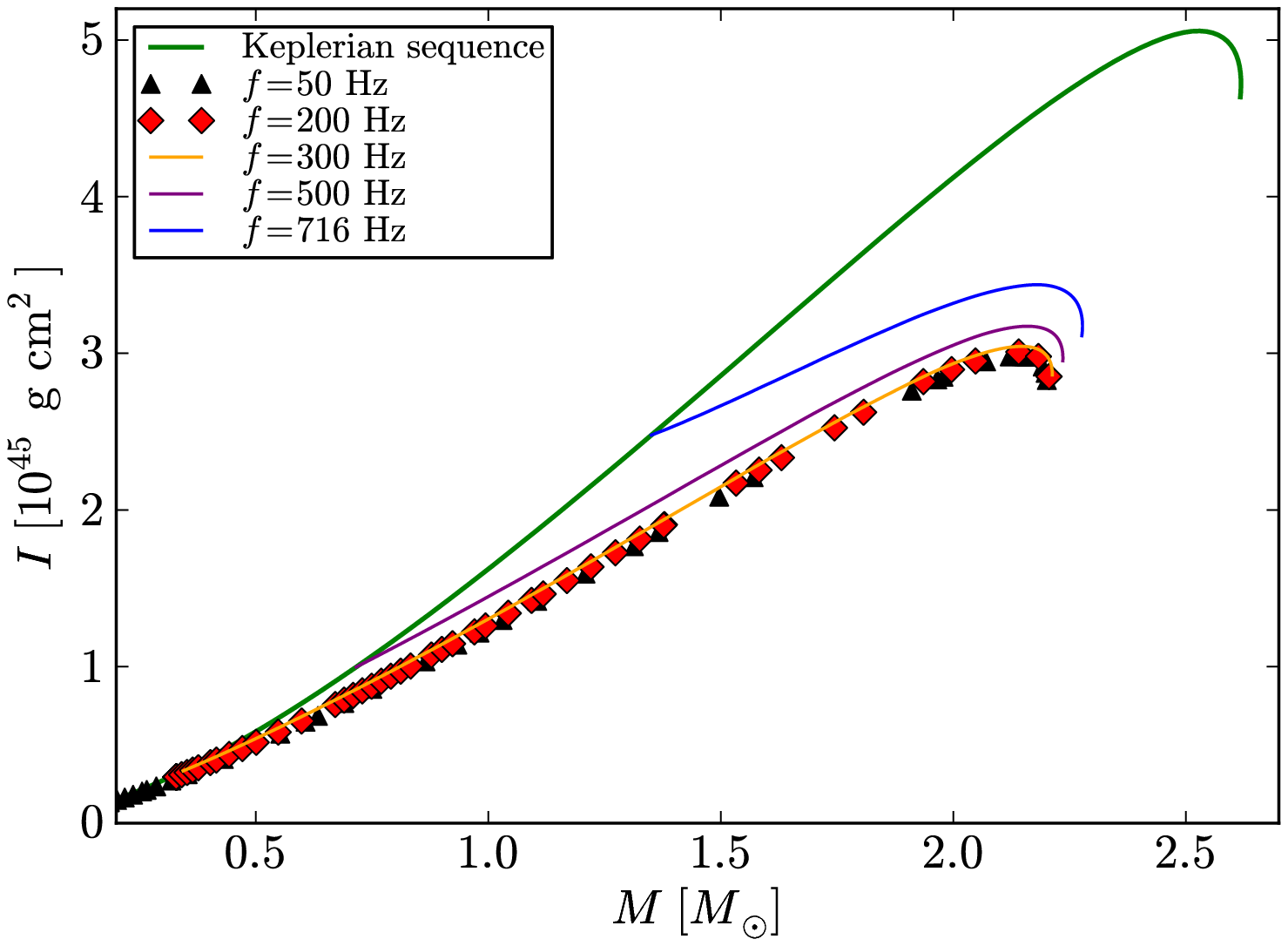}
\includegraphics[width=0.32\hsize,clip]{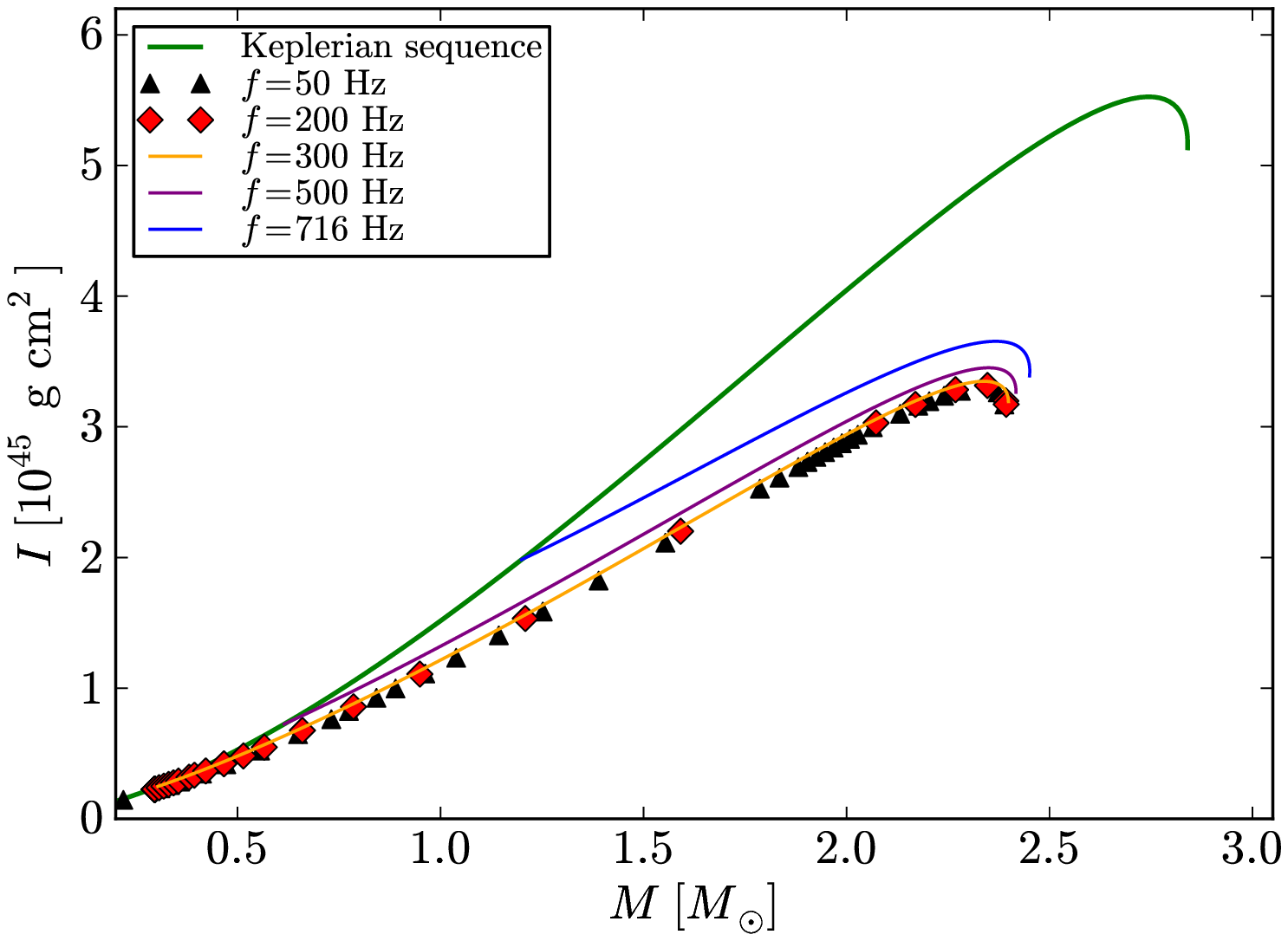}
\includegraphics[width=0.32\hsize,clip]{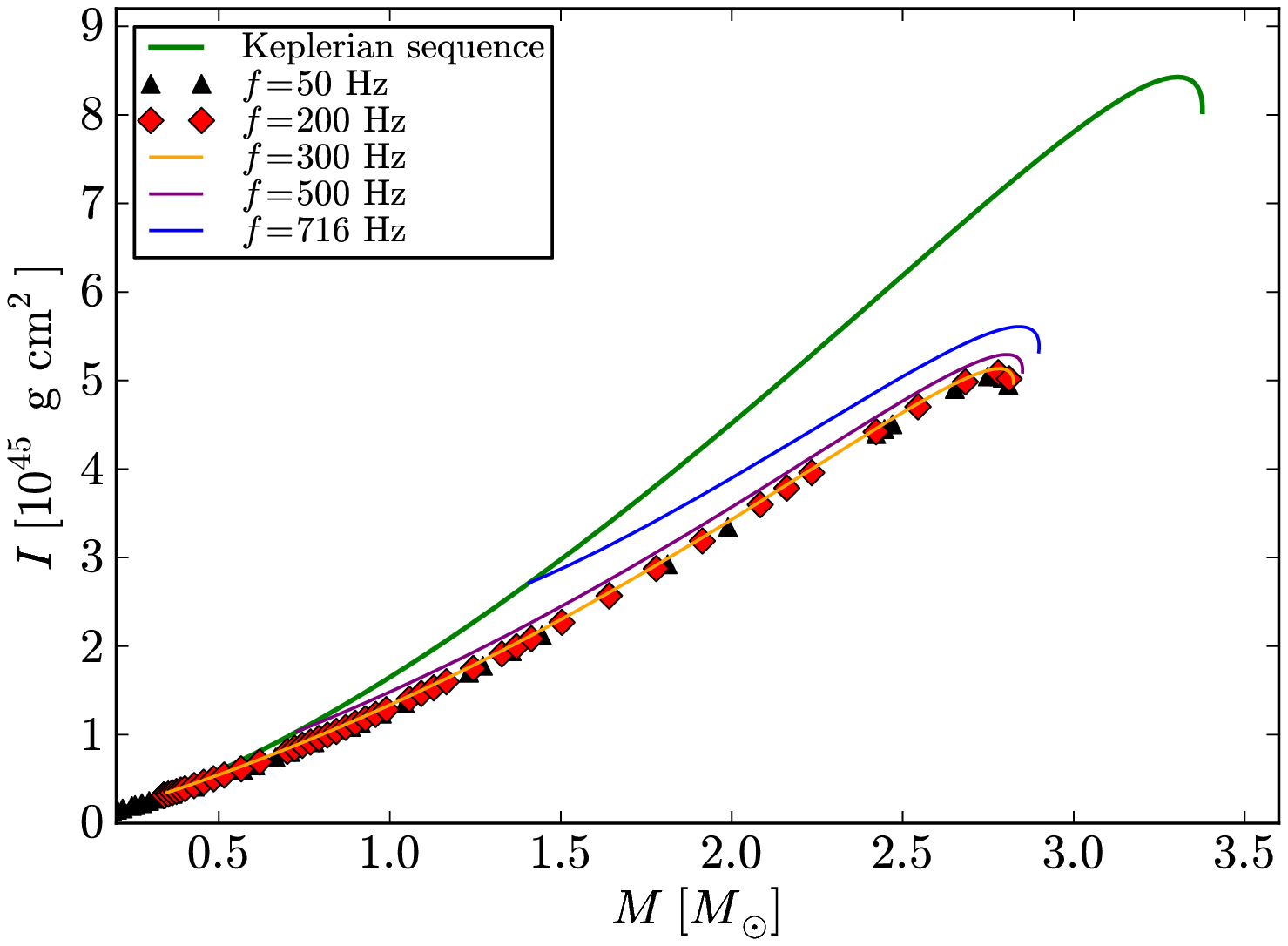}
\caption{\label{fig10}Moment of inertia versus mass relation using EOS TM1, GM1, and NL3 (from top to bottom) set of parameters for same sequences as in figure~\ref{fig03}.}
\end{figure*}

\begin{figure*}[!hbtp]
\includegraphics[width=0.32\hsize,clip]{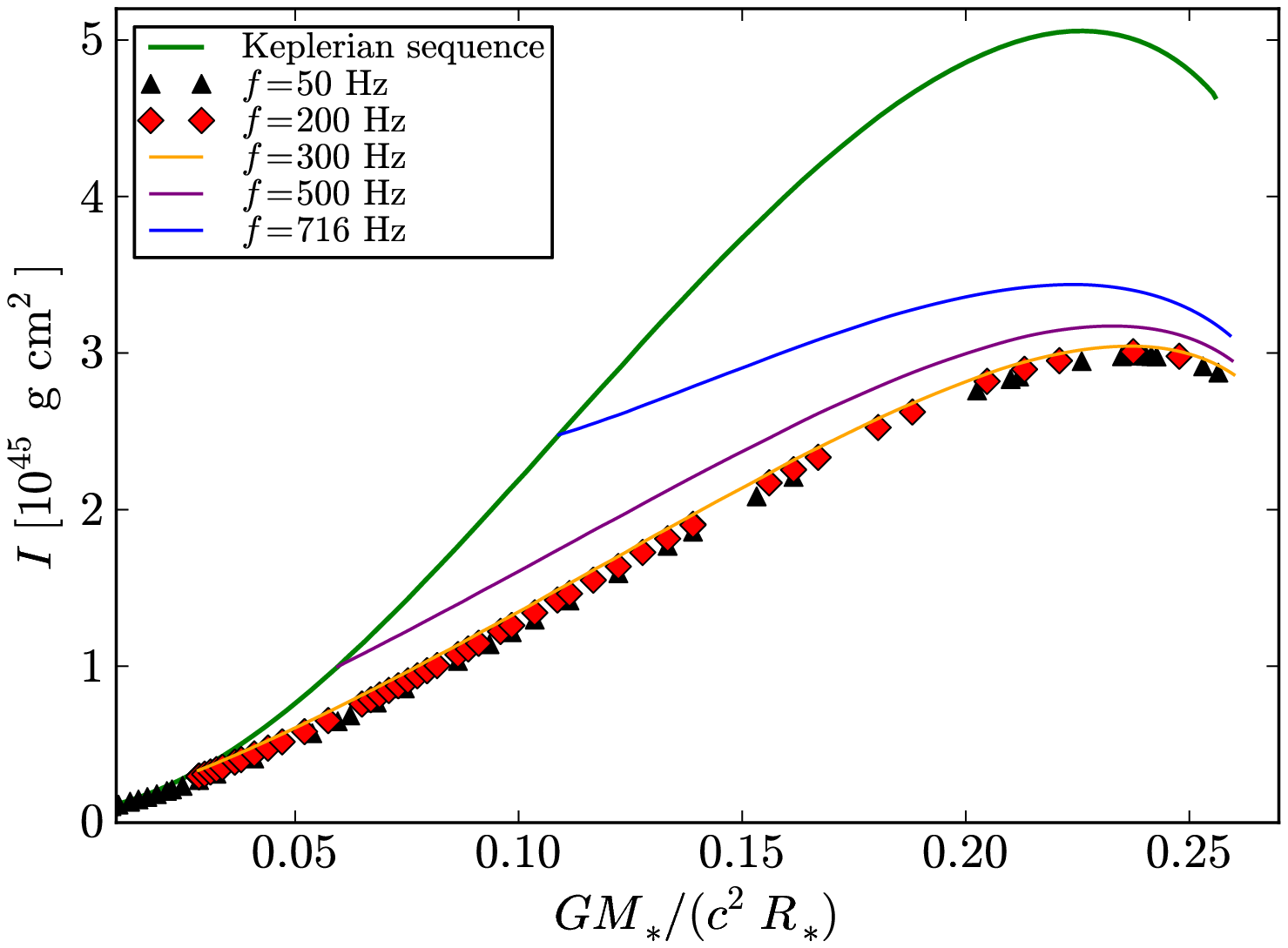}
\includegraphics[width=0.32\hsize,clip]{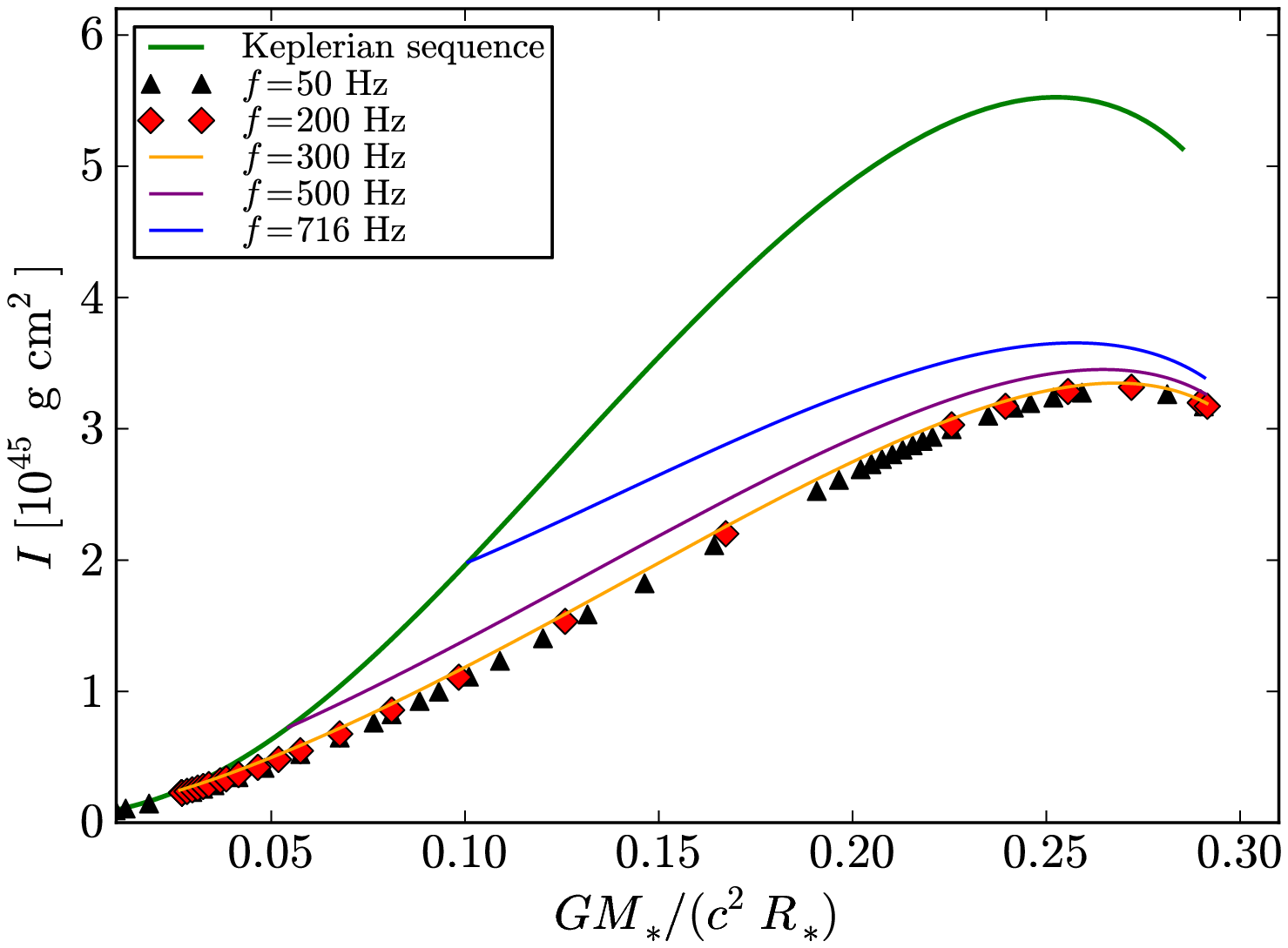}
\includegraphics[width=0.32\hsize,clip]{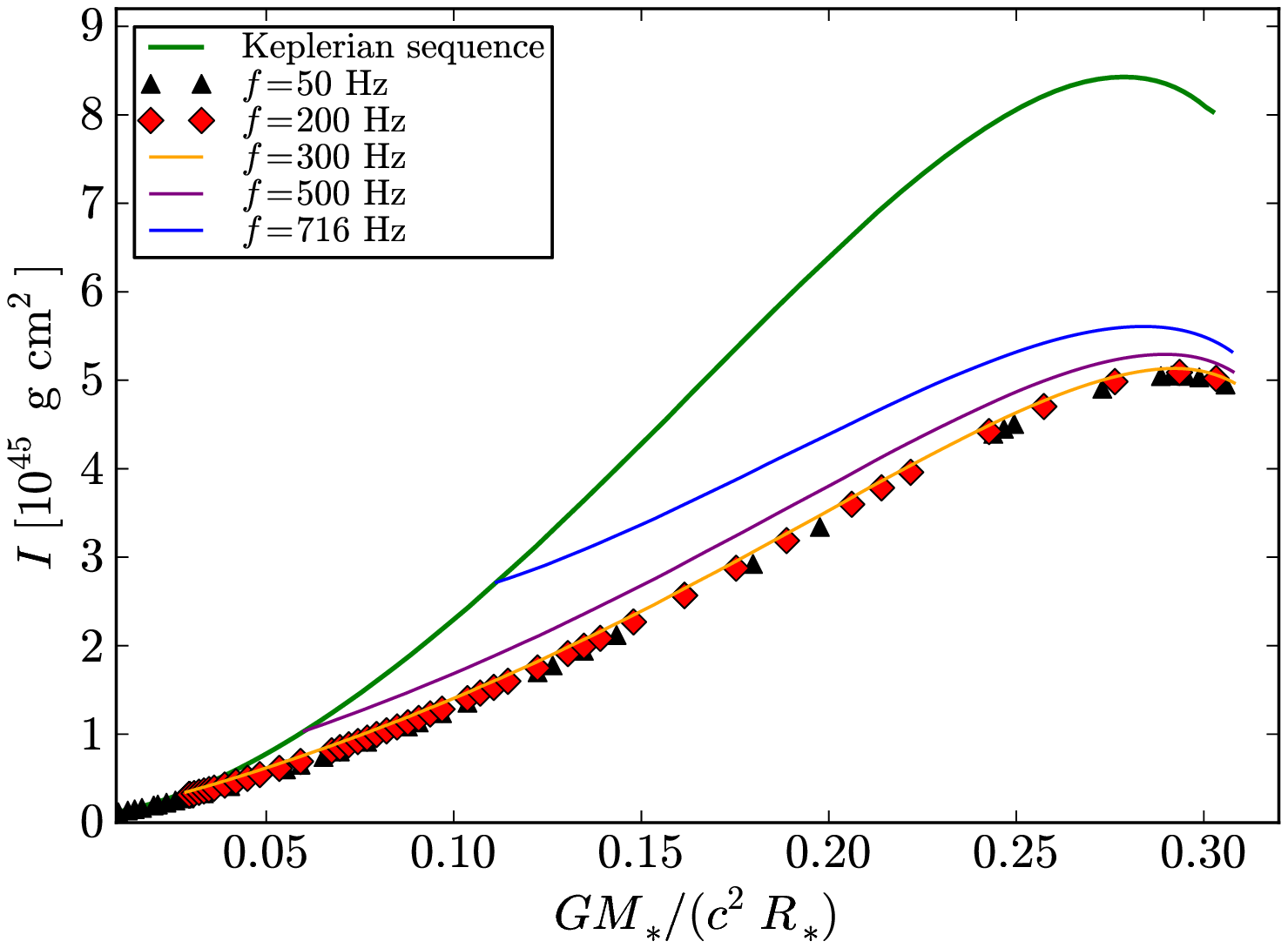}
\caption{\label{fig12}Moment of Inertia versus compactness using EOS with  TM1, GM1, and NL3 (from left to right) set of parameters for same sequences as in figure~\ref{fig03}.}
\end{figure*}

In figure~\ref{fig10} we plot the moment of inertia as a function of the mass for some $\Omega$-constant sequences together with the Keplerian sequence, while in figure~\ref{fig12} we show the relations between $I$ and the compactness, $G M_{\ast}/(c^2 R_{\ast})$, where $M_{\ast}$ and $R_{\ast}$ are the mass and  the radius of the spherical configuration with the same central density as the rotating one, $\varepsilon_{c}$.

The above figures confirm that for rotation frequencies $\lesssim 200$~Hz, or rotation periods $\gtrsim 5$~ms, the deformation of the star is very little and, indeed, the non-rotating or the slow rotation regimes can be safely adopted as accurate approximations of the rotating NS.

\section{Quadrupole Moment}\label{sec:7}

The quadrupole moment in the RNS code is given by
\begin{equation}
\label{eq14}
M_2 = \frac{1}{2} {r_{eq}}^3 \int_{0}^{1} \frac{{s'}^2 d{s'}}{(1-s')^4} \int_{0}^{1} P_{2} (\mu') \tilde{S}_{\rho} (s',\mu') d \mu',
\end{equation} 
where $r_{eq}$ is the value of the coordinate radius at equator, $\rho \equiv 2 \nu - \ln (B)$, $s=r/(r+r_{eq}) \in [0,1]$ is a compacted radial coordinate, $\mu = \cos (\theta)$, $P_{2} (\mu)$ is the Legendre polynomial of second order, and $\tilde{S}_{\rho} = r^2 S_{\rho} $, being $S_{\rho}$ a source function defined as
\begin{widetext}
\begin{eqnarray}
\label{eq15}
S_{\rho} (r,\mu) = &  &  e^{\frac{\gamma}{2}} \Biggl[ 8 \pi e^{2 \lambda} (\varepsilon + P) \frac{1+u^2}{1-u^2} + r^2 e^{-2 \rho} \left[ {\omega}_{,r}^2 + \frac{1}{r^2} (1-{\mu}^2)  {\omega}_{,\mu}^2 \right] + \frac{1}{r} {\gamma}_{,r} - \frac{1}{r^2} \mu {\gamma}_{,\mu}  \nonumber \\
                             & + & \frac{\rho}{2} \left\lbrace 16 \pi e^{2 \lambda} - {\gamma}_{,r} \left( \frac{1}{2} {\gamma}_{,r} + \frac{1}{r} \right) \frac{1}{r^2} {\gamma}_{,\mu} \left[ \frac{1}{2} {\gamma}_{,\mu} (1-{\mu}^2) -\mu \right] \right\rbrace \Biggr],
\end{eqnarray}
\end{widetext}
with $\gamma = \ln (B)$. However, as shown in Ref.~\cite{2012PhRvL.108w1104P}, equation~(\ref{eq14}) is not the actual quadrupole moment of the rotating source according to the Geroch-Hansen multipole moments \cite{1970JMP....11.1955G,1970JMP....11.2580G,1974JMP....15...46H}. Indeed, the quadrupole moment extracted via the Ryan's expansion method \cite{1995PhRvD..52.5707R} is \cite{2012PhRvL.108w1104P,2014PhRvD..89l4013Y}
\begin{eqnarray}
\label{eq17}
M_2^{\rm corr} &=& M_2- \frac{4}{3} \left( \frac{1}{4} + b_0 \right) M^3,\\
b_0 &=& -  \frac{16 \sqrt{2\pi} {r_{eq}}^4}{M^2} \int_0^{\frac{1}{2}} \frac{{s'}^3 ds'}{(1-s')^5}\nonumber \\
          & \times & \int_0^1 d\mu' \sqrt{1-{\mu'}^2} P(s',\mu') e^{\gamma + 2\lambda} T_0^{\frac{1}{2}}(\mu'),\label{eq18}
\end{eqnarray}
where $M_2$ is given by equation~(\ref{eq14}) and $T_0^{\frac{1}{2}}$ is the Gegenbauer polynomial of order $0$ with normalization $T_0^{1/2} = \sqrt{2/\pi} C_0$, with $C_0$ the traditional 0th-order Gegenbauer polynomial.

\begin{figure*}[!hbtp]
\includegraphics[width=0.32\hsize,clip]{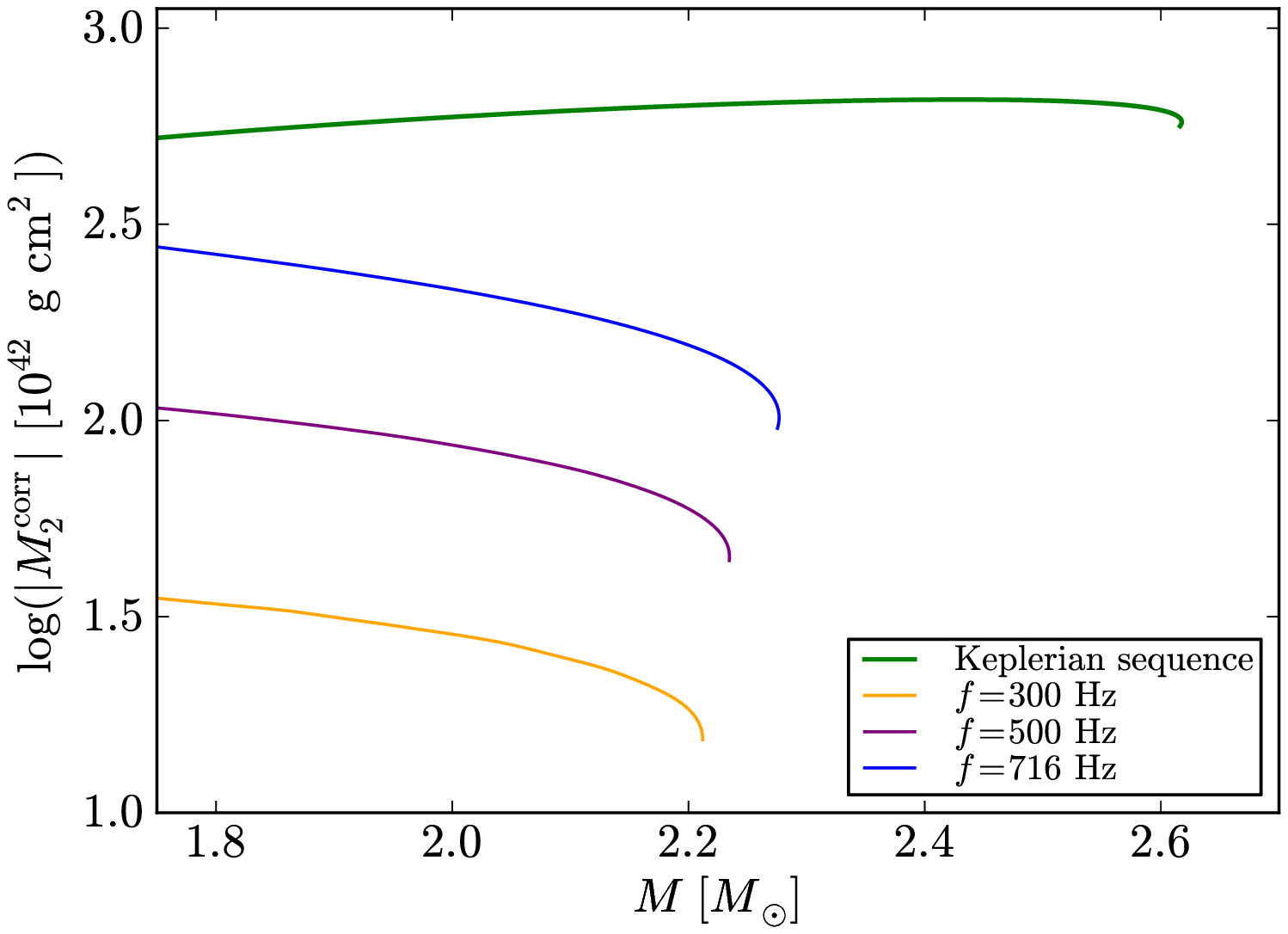}
\includegraphics[width=0.32\hsize,clip]{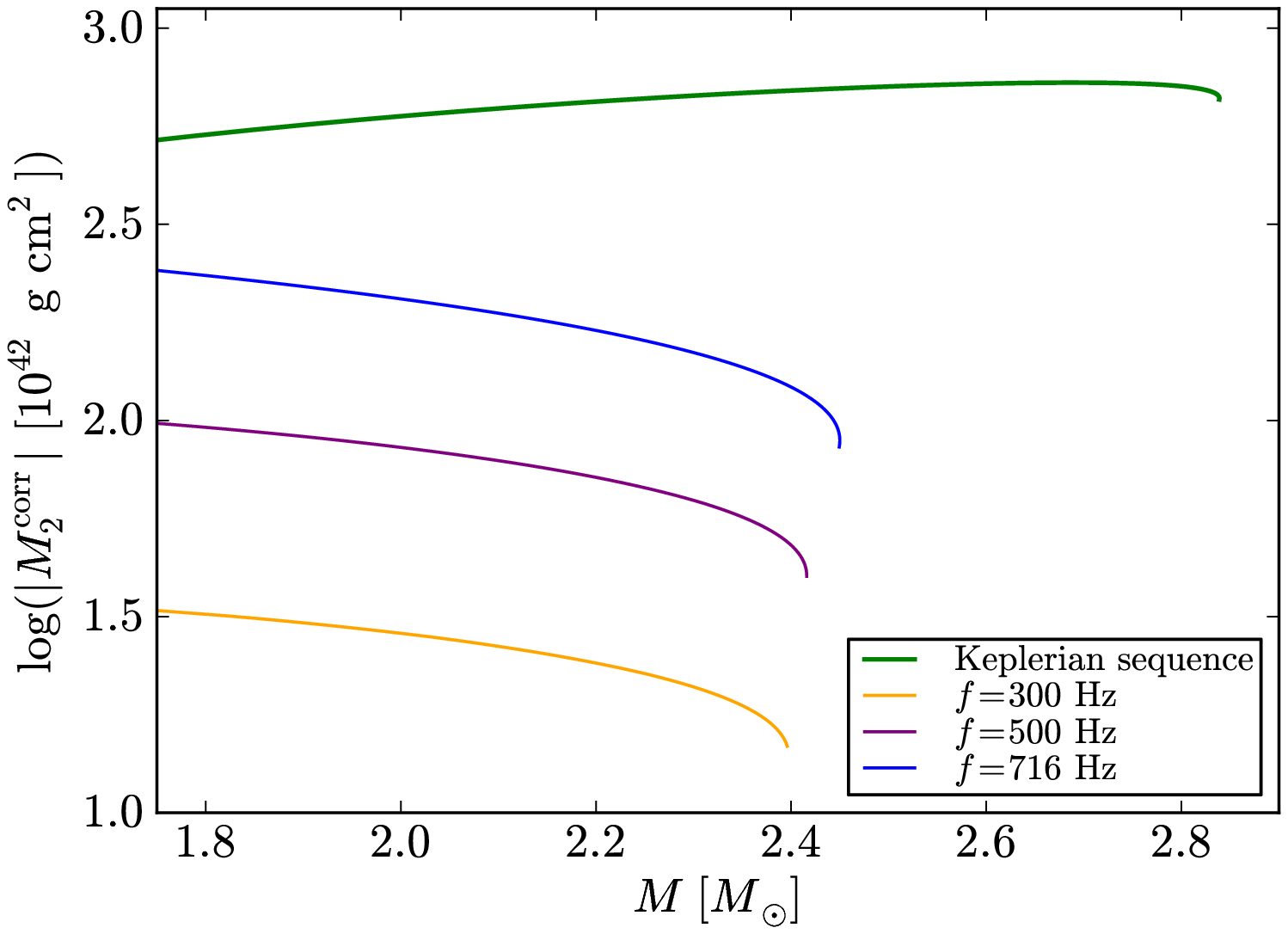}
\includegraphics[width=0.32\hsize,clip]{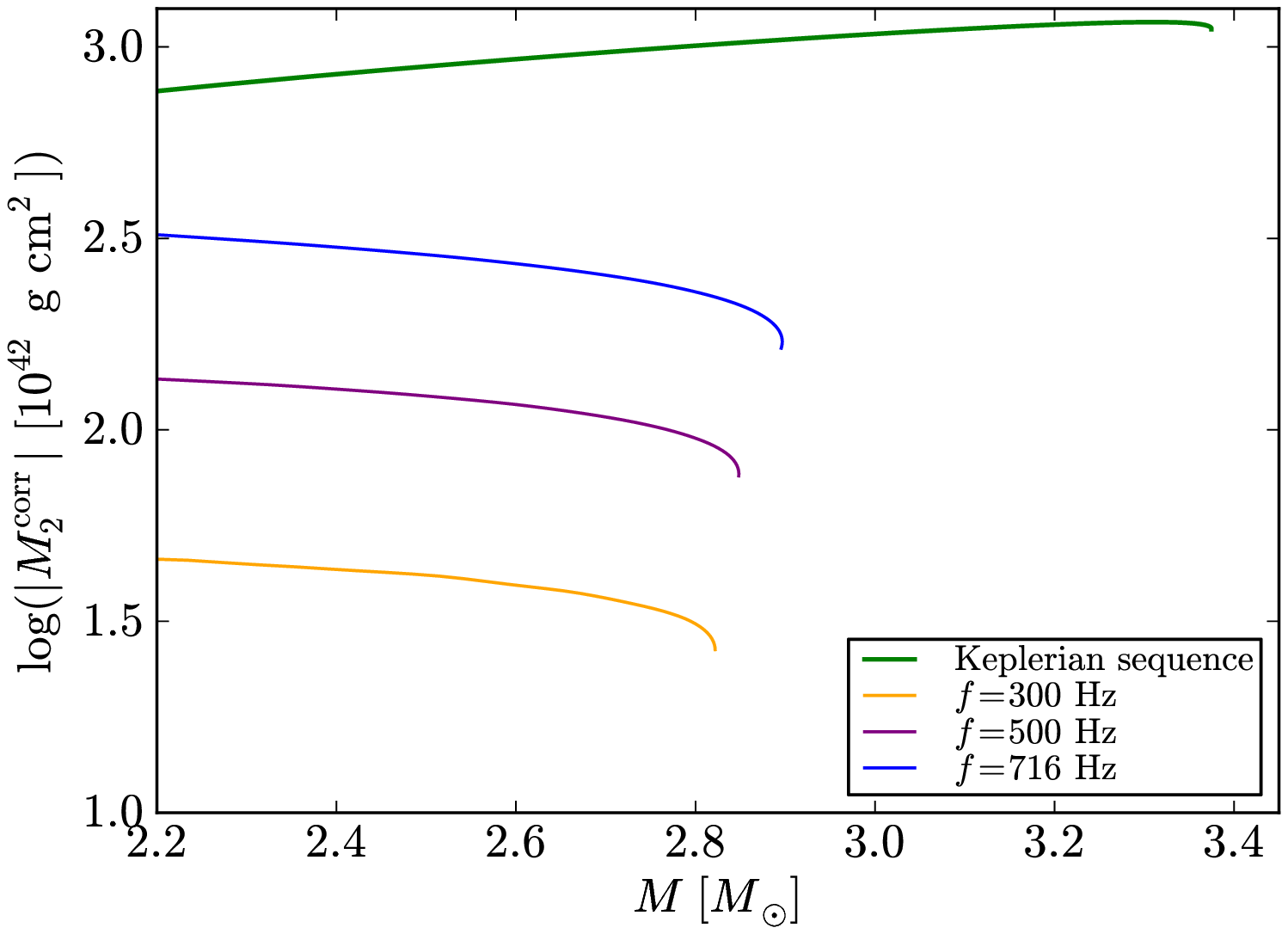}
\caption{\label{fig15}The modulus of corrected value for the mass quadrupole (in logarithmic scale) obtained via equation~(\ref{eq17}) is plotted against gravitational mass for the same constant frequency sequences of figure~\ref{fig03}.}
\end{figure*}

Following Refs.~\cite{2012PhRvL.108w1104P,2014PhRvD..89l4013Y}, we computed numerically the correcting factor $b_0$ given by equation~(\ref{eq18}), and then obtained the corrected quadrupole moment through equation~(\ref{eq17}). In figure~\ref{fig15} the modulus of $M_2^{\rm corr}$ is plotted in logarithmic scale against the gravitational mass for selected constant frequency sequences. Each sequence was stopped at the secular instability limit. We can see that the quadrupole moment is a decreasing function of the mass along a constant frequency sequence while it is an increasing function along the Keplerian sequence.

We turn to compare and contrast the above mass quadrupole moment with the one of the Kerr solution, 
\begin{equation}
\label{eq16}
M_2^{\rm Kerr} = \frac{J^2}{M}.
\end{equation}
The reason for this is twofold. First, we point out the longly discussed question in astrophysics if the Kerr solution may describe the exterior gravitational field of a realistic astrophysical source besides a black hole, namely is there any matter content which could generate a Kerr exterior field? (see, e.g., Refs.~\cite{2006PhRvD..73j4038P,2012PhRvD..86f4043B}, and references therein). Second, if the answer to the previous question is negative, then one can distinguish a NS from a black hole with the same mass and angular momentum from the knowledge of the quadrupole moment (see, e.g., Ref.~\cite{2012ApJ...756...82P}, and references therein).

\begin{figure*}[!hbtp]
\includegraphics[width=0.32\hsize,clip]{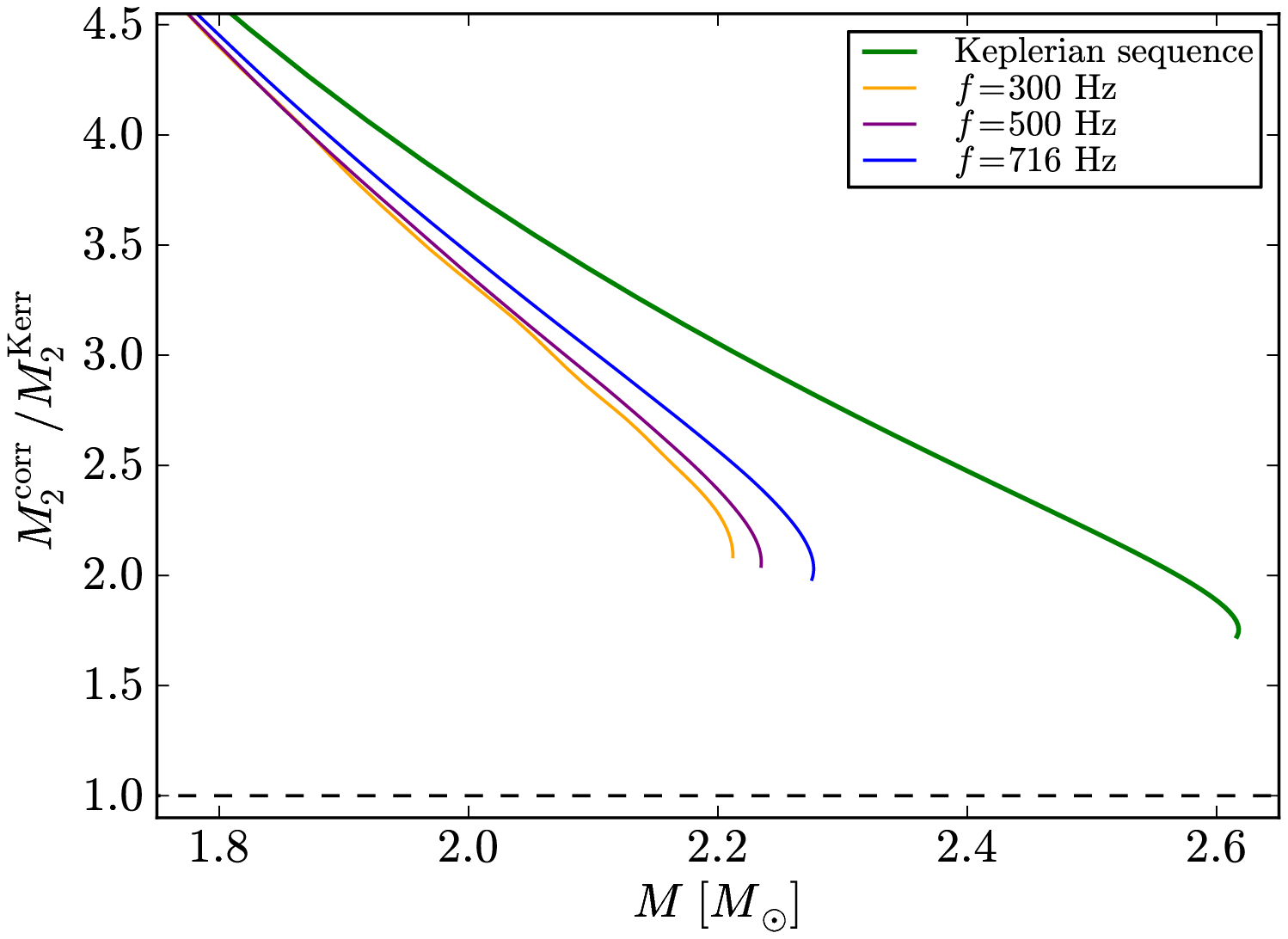}
\includegraphics[width=0.32\hsize,clip]{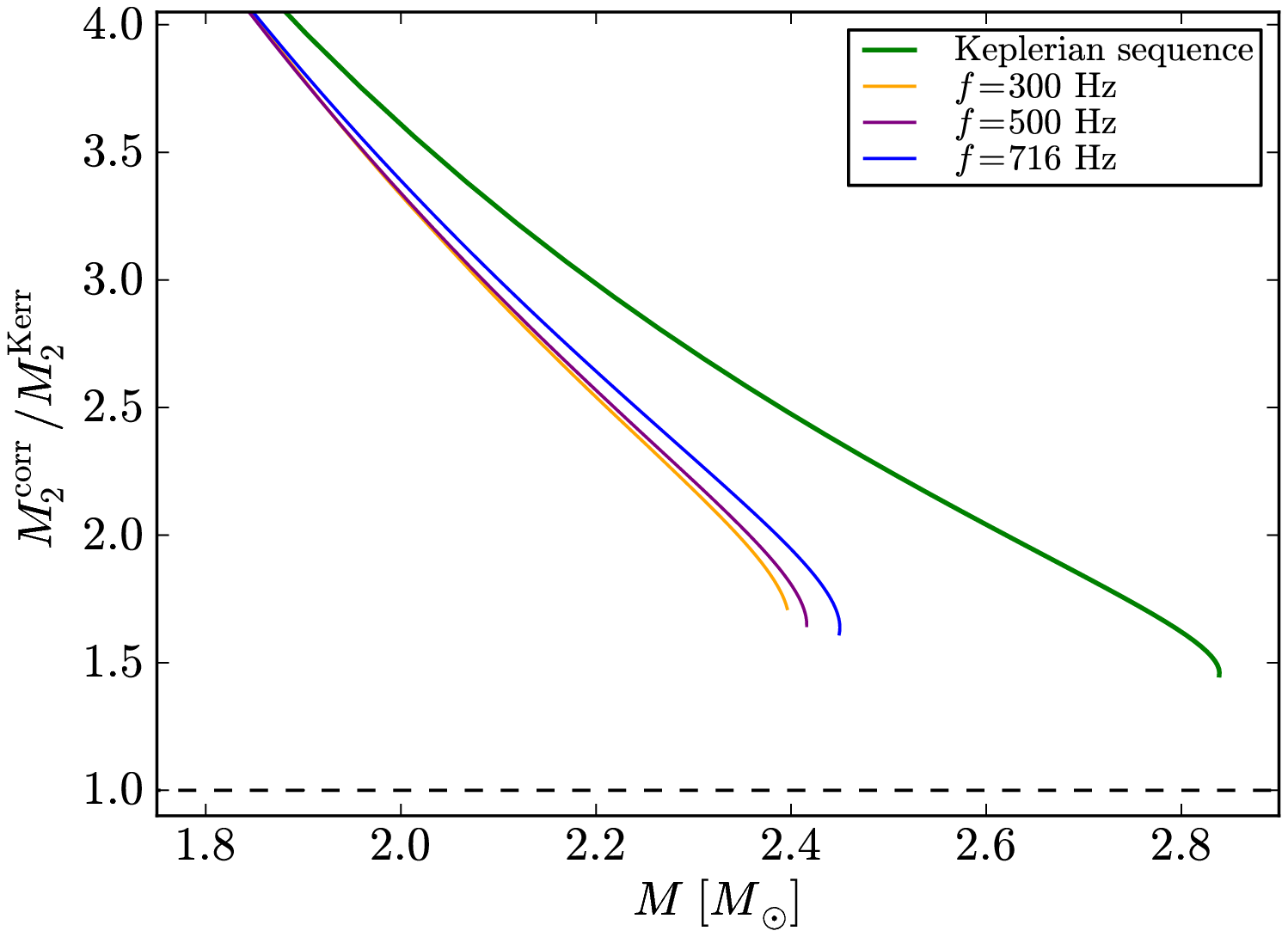}
\includegraphics[width=0.32\hsize,clip]{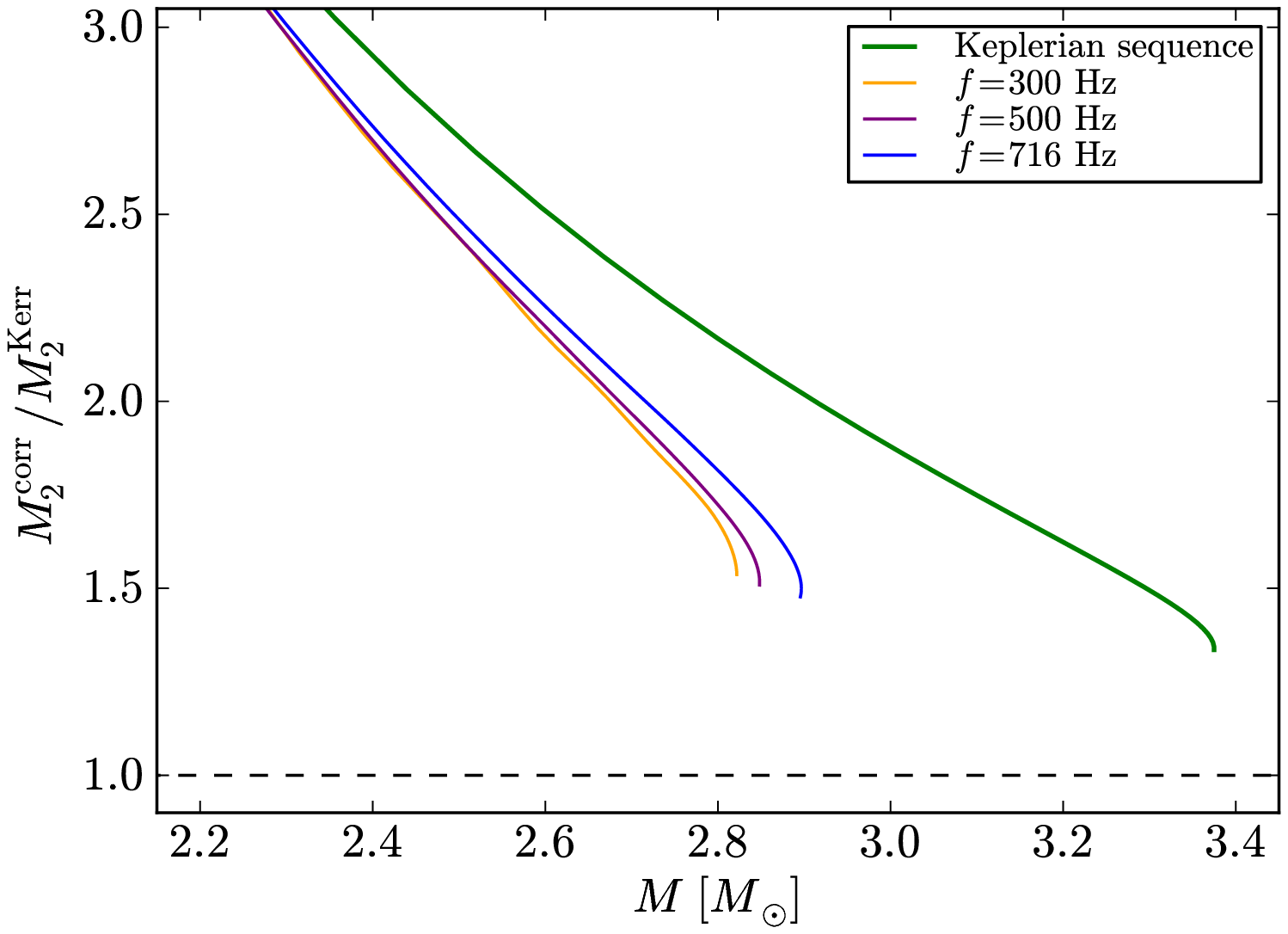}
\caption{\label{fig14} $M_2^{\rm corr}/M_2^{\rm Kerr}$ ratio for the same selected sequences of constant frequency of figure~\ref{fig15} and EOS TM1, GM1, and NL3 (from left to right). We show here only the region of large masses where $M_2^{\rm corr}$ starts to approach the Kerr value $M_2^{\rm Kerr}$.}
\end{figure*}

In figure~\ref{fig14} we show the ratio between the NS quadrupole moment, $M_2^{\rm corr}$ given by equation~(\ref{eq17}), and the Kerr solution quadrupole moment, $M_2^{\rm Kerr}$, for selected constant frequency sequences. We find that $M_2^{\rm corr}$ starts to approach $M_2^{\rm Kerr}$, as intuitively expected, for masses close to the maximum stable value. An interesting feature that we can see from figure~\ref{fig14} is that the stiffer the EOS the more the quadrupole moment approaches the Kerr value. This result is well in accordance with previous results that showed that the compactness of the star increases, also the moment of inertia, Love numbers and mass quadrupole approach the ones of a black hole, though they will never coincide (see, e.g., Ref.~\cite{2013PhRvD..88b3009Y}). Moreover, we confirm in the full rotation regime, the previous result obtained in the slow-rotation Hartle's approximation \cite{2013MNRAS.433.1903U}, that the ratio $M_2^{\rm corr}/M_2^{\rm Kerr}$ is a decreasing function of the NS mass, hence reaching its lowest value at the maximum mass configuration. Indeed, as we can see from figure~\ref{fig14}, the largest the maximum mass attained by a NS model, the closest the NS quadrupole moment approaches the Kerr solution value, reaching even values $<1.5$ for stiff EOS such as the NL3 model.

\section{Discussions and Conclusions}\label{sec:8}

We have computed uniformly rotating NSs for selected relativistic mean-field nuclear matter model EOS (TM1, GM1, and NL3). Specifically, we have calculated their gravitational mass, equatorial and polar radii, eccentricity, angular momentum, moment of inertia and quadrupole moment. We have established the region of stability against mass-shedding and the secular axisymmetric instability. We have provided plots of all these physical quantities e.g. as a function of the mass of the configurations. We have also constructed sequences of constant rotation frequency and determined approximately the rotation rate at which deviations of the structure parameters from the spherically symmetric (or slowly rotating) values start, obtaining $f\approx 200$~Hz, a value in agreement with previous works (see, e.g., Ref.~\cite{benhar05}). 

From the astrophysical point of view, we have obtained a lower bound for the mass of the fastest observed pulsar, PSR J1748--2446ad with $f=716$~Hz, by constructing its constant rotation frequency sequence and constraining it to be within the stability region: we obtained $M_{\rm min}=[1.2$--$1.4]~M_\odot$, for the EOS we have used in this work, a prediction submitted for observational verification. We have obtained also a fitting formula relating the baryonic and gravitational mass of non-rotating NSs [see equation (\ref{eq:MbM})] independent on the EOS. We have computed a formula for the masses of NSs on the secular instability line as a function of their angular momentum, see equation (\ref{eq:secformula}). We studied the Kerr parameter (dimensionless angular momentum) of NSs and found that it reaches a maximum value $(a/M)_{\rm max}\approx 0.7$, independent on the EOS. This result bring us to the important conclusion that the gravitational collapse of a uniformly rotating NS, constrained to mass-energy and angular momentum conservation, cannot lead to a maximally rotating Kerr black hole, which by definition has $(a/M)_{\rm BH,max}= 1$. We have also shown that the quadrupole moment of realistic NSs does not reach the Kerr value (for the same values of mass and angular momentum), but this is closely approached from above at the maximum mass value, in physical agreement with the no-hair theorem. We have also found that the stiffer the EOS the closer the Kerr solution is approached.

It is important to stress that the results shown in this work for some specific nuclear EOS likely will remain valid in the case of other different models, providing they are consistent with current observational constraints, especially the mass of PSR J0348+0432, $M=2.01 \pm 0.04 M_\odot$ \citep{antoniadis13}. The existence of such a massive NS, clearly favors stiff nuclear EOS as the ones obtained via RMF theory, which leads to a critical NS mass higher than this constraint.

To conclude, as already mentioned in sec. \ref{sec:3}, we would like remind the importance of considering a global charge neutrality condition for the system, instead of a local one, which needs a new and more complete code to treat these kind of problems, including the case of fast rotating strange quark stars with crust, which show similar features in the core-crust transition.

\begin{acknowledgements}
C. C. and S. F. would like to acknowledge GNFM-INdAM and ICRANet for partial support. It is a pleasure to thank D. P. Menezes and R.C.R. de Lima for discussions on the equation of state of NSs and for supplying the EOS tables. J.A.R. acknowledges the support by the International Cooperation Program CAPES-ICRANet financed by CAPES -- Brazilian Federal Agency for Support and Evaluation of Graduate Education within the Ministry of Education of Brazil.
\end{acknowledgements}

\end{document}